\documentclass[a4paper,11pt]{article}
\usepackage{jcappub}
\usepackage{amsmath,amssymb}
\usepackage{graphicx}
\usepackage{fontawesome5}
\usepackage{booktabs}
\usepackage{array}
\newcolumntype{L}[1]{>{\raggedright\arraybackslash}p{#1}}
\usepackage{xcolor}
\usepackage{threeparttable}
\usepackage{tikz}
\usepackage[section]{placeins}   %

\usetikzlibrary{positioning,arrows.meta}

\newcommand{\OGal}{h^2\Omega_{\rm Gal}}
\newcommand{\OExt}{h^2\Omega_{\rm Ext}}
\newcommand{\Ob}{h^2\Omega_{\rm b}}
\newcommand{\fb}{f_{\rm b}}
\newcommand{\aone}{a_1}
\newcommand{\Dbar}{\bar{D}}
\newcommand{\Nd}{N_{\rm d}}
\newcommand{\Neff}{N_{\rm eff}}
\newcommand{\Aacc}{A_{\rm acc}}
\newcommand{\Poms}{P_{\rm oms}}
\newcommand{\khat}{\hat{k}}
\newcommand{\shat}{\hat{s}}
\newcommand{\nhat}{\hat{n}}
\newcommand{\lnL}{\ln\mathcal{L}}

\graphicspath{{figures/}}

\title{\boldmath Neural Separation and SBI-Guided Inference of LISA Gravitational-Wave Backgrounds}

\author[a]{Moslem Ahmadvand}
\affiliation[a]{Faculty of Physics, Shahrood University of Technology, P.O.Box 3619995161
	Shahrood, Iran}
\emailAdd{ahmadvand@ipm.ir}	
\author[b]{Saeed Ansarifard}
\affiliation[b]{School of Physics, Institute for Research in Fundamental Sciences (IPM), P.O. Box 19395-5531, Tehran, Iran}
\emailAdd{ansarifard@ipm.ir}

\abstract{A first-order phase transition in the early Universe could produce a
gravitational-wave background detectable by LISA at millihertz frequencies. Recovering this background requires distinguishing it
from instrument noise and astrophysical foregrounds. We assess the role of machine learning using
simulated four-year LISA data, for which the exact likelihood is known in
closed form and serves as the reference. A neural network first separates
instrument noise from the combined gravitational-wave signal. Thirty-two
principal components of the separated spectra give signal-parameter
uncertainties comparable to those from about 300 principal components of the
raw spectra.
A second network, trained by simulation-based inference (SBI), estimates
parameter posteriors with reliable uncertainties 3--6 times wider
than exact. Retrained on simulations concentrated around a single observation,
it narrows this gap to a factor of 1.4--2.8.
When the fit has several competing solutions, however, the exact likelihood is
still needed to determine their relative probabilities. Guiding the
exact likelihood with the SBI posterior closely approximates the reference posterior, with
interval widths at most about 15\% too narrow on a benchmark simulation, using
15--60 times fewer likelihood evaluations than Markov-chain Monte Carlo. This
guided analysis passes a calibration test fixed in advance on 100 unseen
simulations.
}


\begin{document}
\emergencystretch=1.5em
\maketitle
\flushbottom

\section{Introduction}\label{sec:intro}

The Laser Interferometer Space Antenna (LISA)~\cite{AmaroSeoane:2017,Colpi:2024}
will measure gravitational waves between approximately $10^{-4}$ and
$10^{-1}$\,Hz. A first-order phase transition in the early Universe, occurring
near the electroweak scale, could have produced a stochastic
gravitational-wave (GW) background detectable in this band. Such a transition
proceeds through the nucleation, expansion and collision of bubbles of the new
phase. The resulting anisotropic stresses, generated by the bubble walls and
the plasma motions, source GWs~\cite{Caprini:2016,Caprini:2020}. Today, such a
background would appear as an energy-density spectrum $\Omega_{\rm GW}(f)$,
which can exhibit a characteristic peak. The peak frequency is determined
primarily by the transition temperature and duration, while its amplitude
depends on the transition strength and duration. Recovering this spectrum, from
which the thermodynamic parameters of the transition can be inferred, is one of
LISA's cosmological goals~\cite{Caprini:2024,Guan:2025}.

The cosmological background is not the only signal in LISA's frequency band.
Most of the millions of double-white-dwarf binaries in our Galaxy are too
numerous and too closely spaced in frequency to be resolved individually, and
together they form a stochastic foreground that can exceed the instrument noise
near a millihertz, where a phase-transition signal may also peak. Because
LISA's sensitivity depends on source direction, the observed strength of the
Galactic double-white-dwarf foreground varies through the year as the
constellation changes orientation along its
orbit~\cite{Boileau:2021,Hindmarsh:2024}. A second, weaker foreground comes
from extragalactic binaries containing stellar-mass black holes or neutron
stars~\cite{Karnesis:2021}. These foregrounds add to the instrument's internal
noise~\cite{Adams:2013qma}. Unlike ground-based observatories, which can suppress noise by
cross-correlating independent detectors~\cite{Allen:1999}, LISA cannot use this
technique because it is a single observatory. Instead, it can monitor
instrument noise using a data channel that is nearly insensitive to
GWs~\cite{Guan:2025,Adams:2010}.

The European Space Agency adopted LISA in 2024~\cite{ESA:LISA2024}, with launch
planned for the mid-2030s and a nominal science mission of at least 4.5
years~\cite{Colpi:2024,Caprini:2024}. Until
data arrive, analysis methods are developed and tested on simulated data, in
which the injected signals and their parameters are known. Stochastic
backgrounds are typically analysed through the power spectral density (PSD) of
the data, by comparing the observed spectrum with models of the signal and
noise. For stationary Gaussian noise, the likelihood has a closed form and Markov-chain Monte Carlo
(MCMC) can sample the resulting
posterior~\cite{Caprini:2024,Guan:2025,Flauger:2021}. Such template-based
methods have also been used to connect GW spectra to
particle-physics models of the phase transition~\cite{Guan:2025,Liang:2025}.
They provide exact inference within the assumed model. However, real LISA data will
also contain noise transients (glitches) and overlapping signals,
which make the likelihood difficult to formulate~\cite{Houba:2025}. Moreover,
MCMC requires many likelihood evaluations, and when signal components overlap
in frequency the posterior can have several competing solutions that MCMC
explores slowly or misses.

Simulation-based inference (SBI)~\cite{Cranmer:2020} avoids the need for an
explicit likelihood. Instead, a neural network learns the posterior from
simulated data generated with known parameters. Once trained, it returns the
posterior for new data at negligible cost, without evaluating the likelihood, and it
can work from summaries of the data, including the outputs of other neural
networks. Its uncertainties can also be tested on independent simulations with
known parameters. Any effect that can be simulated, such as glitches or
overlapping signals, is therefore accounted for without being modelled in a
likelihood. For example, Ref.~\cite{Alvey:2023} uses SBI to reproduce MCMC
results for 12 days of mock LISA data, modelling a single stochastic background
and two noise amplitudes in one data channel, with the network updated for
each observation. Ref.~\cite{Dimitriou:2023knw} uses likelihood-free inference
to reconstruct the spectral shape of a background observed by LISA. Neural
networks have also been used to separate stochastic
backgrounds from detector noise in ground-based
observations~\cite{Einsle:2025,Einsle:2026} and to separate overlapping LISA
signals and glitches~\cite{Houba:2025}.

\begin{figure}[tbp]
\centering
\begin{tikzpicture}[
  font=\footnotesize, >={Stealth[length=2mm]},
  box/.style={draw, rounded corners=2pt, align=center, inner sep=3pt, minimum height=8mm},
  result/.style={align=left, anchor=west, font=\footnotesize, text width=48mm}]
\node[box, text width=15mm] (sim) at (0,0) {simulation};
\node[box, text width=18mm] (spec) at (2.55,0) {averaged spectrum in A, E, T};
\node[box, text width=27mm] (mcmc) at (6.05,1.5) {exact-likelihood MCMC};
\node[box, text width=27mm] (net) at (6.05,0) {network separates noise and signal};
\node[box, minimum width=9mm] (sbi) at (8.45,0) {SBI};
\node[box, text width=27mm] (gd) at (6.05,-1.5) {SBI-guided exact inference};
\draw[->] (sim) -- (spec);
\draw[->] (spec.east) -- ++(0.3,0) |- (mcmc.west);
\draw[->] (spec) -- (net);
\draw[->] (spec.east) -- ++(0.3,0) |- (gd.west);
\draw[->] (net) -- (sbi);
\draw[->] (sbi.south) |- (gd.east);
\node[result] at (9.25,1.5) {exact posterior (reference),\\ 72\,000 nominal, 288\,000\\ converged likelihood evaluations};
\node[result] at (9.25,0) {calibrated posterior, 3--6$\times$ wider,\\ no likelihood evaluations after training};
\node[result] at (9.25,-1.5) {near-exact, correctly weighted posterior,\\ 15--60$\times$ fewer evaluations};
\end{tikzpicture}
\caption{The three analyses compared in this paper, all reading the same
averaged spectrum of a simulation (Sec.~\ref{sec:analyses}). The separating network
provides noise and signal spectra for SBI, which gives fast calibrated posteriors, and the exact likelihood,
guided by SBI, gives the final answer.}
\label{fig:pipeline}
\end{figure}
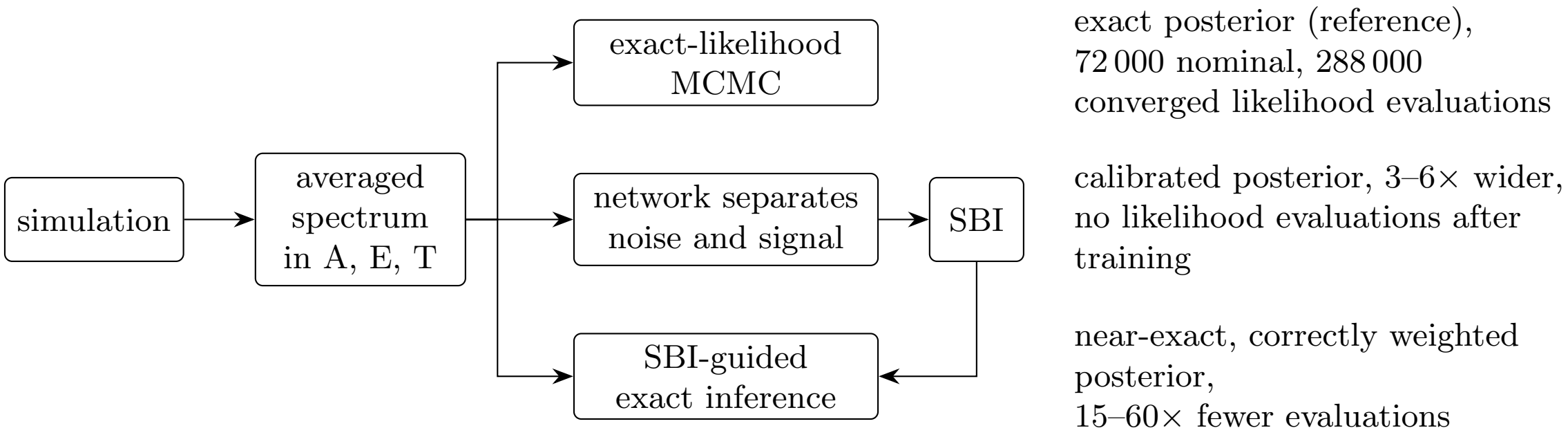

In this paper we simulate four years of LISA data as 126 segments of
$10^{6}$\,s in the three time-delay-interferometry channels A, E and T, of which
T is the noise-monitoring channel, and call one such four-year data set a
\emph{simulation}. Every analysis reads the same input, the PSD of each channel
averaged over the 126 segments, and we compare three analyses of it
(Fig.~\ref{fig:pipeline}). The first is the exact likelihood sampled by MCMC.
Because the simulated noise is stationary and Gaussian and the signal model is
known, this likelihood describes the averaged spectrum exactly, apart from a
small effect of the Galactic modulation on its scatter (App.~\ref{app:gamma}),
the neglected GW signal in T and the excluded frequency windows near the nulls
of T (Sec.~\ref{sec:exact}).
It therefore provides the reference for evaluating the other methods. In this
likelihood, we use the T channel to constrain the instrument noise and neglect
the weak GW signal in T~\cite{Guan:2025}. The second analysis is SBI that reads
the output of a neural network, adapted from
Refs.~\cite{Einsle:2025,Einsle:2026}, which splits each channel's spectrum into
instrument noise and the combined GW signal. The third is an exact-likelihood
analysis guided by the SBI posterior.

To provide a compact input to SBI, we use principal-component
analysis to reduce the separated spectra to 32 coefficients. These are the
weights of the leading 32 combinations of spectral bins that capture most of
the variation across the training simulations. A Fisher-information calculation
indicates that applying the same procedure directly to the measured spectra,
before separation, requires about 300 coefficients to preserve comparable
information about the signal parameters.
The separated spectra also have a physical meaning, as noise and signal
spectra, and are not tied to the parameterisation used for inference. They could therefore in principle be reused with other signal models. The SBI network produces
posteriors at negligible cost once trained and passes all calibration tests set in advance, although
its uncertainties are three to six times wider than those of the exact
likelihood. Retraining on simulations concentrated around a particular
observation (sequential SBI) narrows this gap, reducing the ratio of posterior
widths to 1.4--2.8 for a benchmark simulation.
In 42 of 100 simulations, searches started from different points end at
different likelihood peaks, whose relative probabilities still require the
exact likelihood. Guided by the SBI
posterior, exact inference locates these solutions, flags cases where one may
have been missed, and closely approximates the reference posterior, with
interval widths differing from the reference by 10--15\%
at the benchmark, using 15--60 times fewer
likelihood evaluations than MCMC.
Finally, separating the instrument noise from the signal does not resolve the
degeneracies limiting the phase-transition parameters, because these arise
between the GW signal components rather than between signal
and noise.

Section~\ref{sec:model} describes the simulated data and
Sec.~\ref{sec:analyses} the inference methods. Section~\ref{sec:results} presents
what the separating network retains (Sec.~\ref{sec:role}), the posteriors at a
benchmark point together with the computational cost (Sec.~\ref{sec:benchmark}),
and the reliability of the uncertainties (Sec.~\ref{sec:reliability}).
Section~\ref{sec:conclusions} discusses why separating the noise cannot do
more and gives an outlook.

\section{LISA data model and simulations}\label{sec:model}

\subsection{Detector channels, noise and response}\label{sec:channels}

LISA is a triangle of three spacecraft separated by arms of length
$L=2.5\times10^{9}$\,m. %
Laser light exchanged between the spacecraft measures changes in the arm lengths. However, the laser's own frequency noise is many orders of magnitude larger
than any GW signal. Time-delay interferometry (TDI) suppresses laser frequency
noise by combining measurements taken at suitably delayed
times~\cite{Tinto:2005}. The simplest combinations, $X$, $Y$ and $Z$, each act
like a Michelson interferometer formed by the two arms that meet at one
spacecraft. We take the three arms to be equal and constant, with identical
noise in all spacecraft~\cite{Caprini:2024}. Under this assumption the
combinations
\begin{equation}
  A=\frac{Z-X}{\sqrt2},\qquad E=\frac{X-2Y+Z}{\sqrt6},\qquad T=\frac{X+Y+Z}{\sqrt3},
\end{equation}
have uncorrelated noise. This is the first-generation form of TDI. Real arms
differ and change along the orbit, which makes the channel noises slightly
correlated, a limitation we return to in Sec.~\ref{sec:outlook}. A and E
respond equally to GWs. In T, the symmetric
sum, an isotropic background almost cancels, and T therefore records almost
only the instrument noise~\cite{Guan:2025}.

Two noise sources dominate. Acceleration noise, from spurious forces on the
test masses, dominates at low frequency, and optical-metrology noise, from
errors in measuring the distances between test masses, dominates at high
frequency~\cite{Caprini:2024,Guan:2025,Robson:2019}. As power spectral
densities of fractional frequency fluctuations, they are
\begin{align}
  S_{\rm acc}(f)&=\Aacc^2\,10^{-30}\Big[1+\Big(\frac{0.4\,{\rm mHz}}{f}\Big)^2\Big]
  \Big[1+\Big(\frac{f}{8\,{\rm mHz}}\Big)^4\Big]\frac{1}{(2\pi fc)^2},\\
  S_{\rm oms}(f)&=\Poms^2\,10^{-24}\Big[1+\Big(\frac{2\,{\rm mHz}}{f}\Big)^4\Big]
  \Big(\frac{2\pi f}{c}\Big)^2,
\end{align}
where $c$ is the speed of light. The nominal amplitudes are $\Aacc=3$ (in
fm\,s$^{-2}$\,Hz$^{-1/2}$) and $\Poms=15$ (in pm\,Hz$^{-1/2}$). In each
simulation both are drawn from Gaussian distributions centred on these values
with a 5\% standard deviation. %
Combining the two sources through the TDI transfer functions gives the noise of
each channel~\cite{Guan:2025},
\begin{align}
  N_{\rm A}=N_{\rm E}&=8\sin^2x\,\big[4(1+\cos x+\cos^2x)\,S_{\rm acc}+(2+\cos x)\,S_{\rm oms}\big],\\
  N_{\rm T}&=16\sin^2x\,\big[2(1-\cos x)^2\,S_{\rm acc}+(1-\cos x)\,S_{\rm oms}\big].\footnotemark
\end{align}
\footnotetext{The noise spectra as printed in Ref.~\cite{Caprini:2024},
Eqs.~(A.4)--(A.5), carry an additional factor of the sky-averaged response
$\tilde R_I$ of Eq.~\eqref{eq:response}. We use the form of Refs.~\cite{Guan:2025,Robson:2019}, which
reproduces the LISA sensitivity curve of Ref.~\cite{Robson:2019}.}
where $x\equiv2\pi fL/c$.

\begin{figure}[tbp]
\centering
\includegraphics[width=0.98\textwidth]{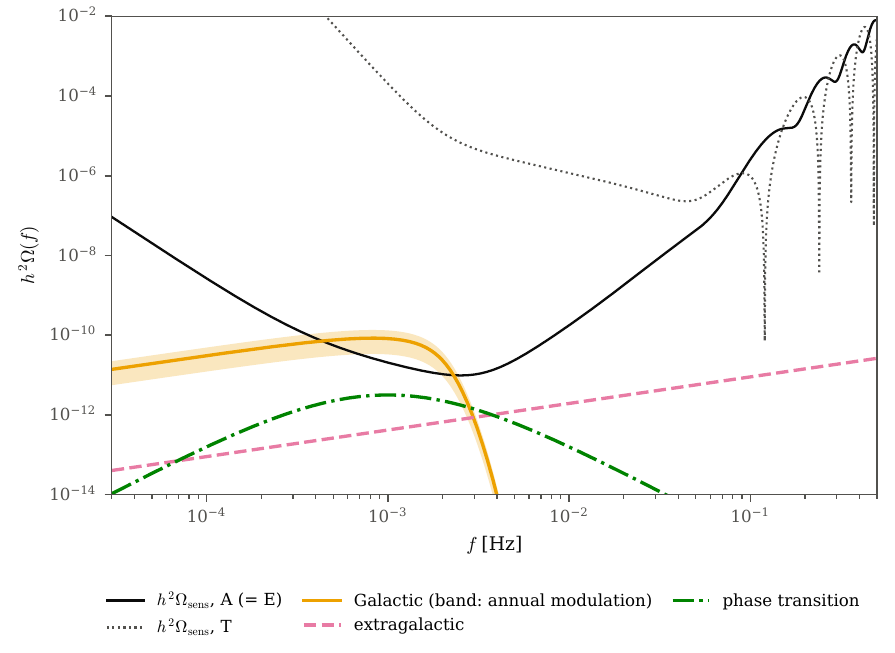}
\caption{Sensitivity of the A (= E) and T channels at nominal noise, expressed
as the background $h^2\Omega_{\rm sens}$ whose signal power would equal the
noise power, together with the three signal components at the benchmark point
(Sec.~\ref{sec:components}). The shaded band is the range of the Galactic
foreground over the year, with $m(t)$ of Eq.~\eqref{eq:modulation} between $1-d_1-d_2$ and $1+d_1+d_2$, at $d_1=0.1$ and
$d_2=0.5$, the centre of the prior in Table~\ref{tab:setup}.}
\label{fig:sensitivity}
\end{figure}

A background with spectrum $\Omega_{\rm GW}(f)$ produces a GW strain PSD
$3H_0^2\,\Omega_{\rm GW}/(4\pi^2f^3)$, where
$H_0=100\,h$\,km\,s$^{-1}$\,Mpc$^{-1}$ is the Hubble constant and $h=0.674$.
The expected PSD of channel $I\in\{{\rm A},{\rm E},{\rm T}\}$ is therefore
\begin{equation}
  S_I(f;\vartheta)=N_I(f;\Aacc,\Poms)+R_I(f)\,\frac{3H_0^2}{4\pi^2f^3}\,
  \frac{h^2\Omega_{\rm GW}(f;\theta)}{h^2},
  \label{eq:Pc}
\end{equation}
where $\theta$ denotes the signal parameters (Sec.~\ref{sec:components}) and
$\vartheta=(\Aacc,\Poms,\theta)$ all parameters. The response $R_I$, which
converts GW strain power into power in channel $I$, uses the sky-averaged fits
of Ref.~\cite{Flauger:2021},
\begin{equation}
  R_I=16\,x^2\sin^2x\,\tilde R_I,\qquad
  \tilde R_{\rm A}=\tilde R_{\rm E}=\frac{9}{20}\,\frac{1}{1+0.7x^2},\qquad
  \tilde R_{\rm T}=\frac{9}{20}\,\frac{x^6}{1.8\times10^{3}+0.7x^8}.
  \label{eq:response}
\end{equation}
The factor $x^6$ in $\tilde R_{\rm T}$ expresses the near-cancellation of the
background in T at low frequency. Figure~\ref{fig:sensitivity} shows the
resulting sensitivity of the A and T channels together with the three signal
components. The sharp dips of the T curve at even multiples of
$c/2L\approx0.060$\,Hz arise because the noise of T vanishes there faster than
its response, a feature of the equal-arm model.

\subsection{Signal components}\label{sec:components}

The Galactic foreground is described by the
fit~\cite{Caprini:2024,Karnesis:2021}
\begin{equation}
  h^2\Omega_{\rm Gal}(f)=\tfrac12\big(f/{\rm Hz}\big)^{2/3}
  \Big[1+\tanh\Big(\frac{f_{\rm knee}-f}{f_2}\Big)\Big]e^{-(f/f_1)^{\nu}}\,\OGal,
\end{equation}
where $\OGal$ without argument is the amplitude we infer. The cut-off frequency
$f_1$ and the knee frequency $f_{\rm knee}$ decrease with the observation time,
as more binaries are resolved and removed. For our four-year observation
$f_1=1.55$\,mHz and $f_{\rm knee}=1.94$\,mHz, with $f_2=0.67$\,mHz and
$\nu=1.56$~\cite{Caprini:2024,Karnesis:2021}, which set the width of the knee
and the sharpness of the cut-off.

Because LISA's sensitivity depends on direction and its orientation changes
along its yearly orbit, the recorded Galactic amplitude varies through the
year~\cite{Boileau:2021,Hindmarsh:2024}. Following Ref.~\cite{Hindmarsh:2024},
we describe this variation by its two leading Fourier harmonics, with periods
of one year and half a year,
\begin{equation}
  h^2\Omega_{\rm Gal}(f,t)=h^2\Omega_{\rm Gal}(f)\,m(t),\;\;
  m(t)=1+d_1\cos(\omega_1t+\phi_1)+d_2\cos(2\omega_1t+\phi_2),\;\;
  \omega_1=2\pi/T_{\rm yr},
  \label{eq:modulation}
\end{equation}
averaged over each segment, where $T_{\rm yr}$ is one sidereal year, $d_1$ and
$d_2$ are the modulation depths and $\phi_1$, $\phi_2$ the phases
(Table~\ref{tab:setup}). For simplicity, we
assume the same $m(t)$ in all three channels, although the modulation depth
generally depends on the channel. When the depths are not known in advance, the
time-varying part constrains the products of the Galactic amplitude and the
modulation depths. A larger Galactic amplitude with weaker modulation can
therefore produce the same time variation. The mean spectrum provides a
separate constraint on the amplitude
(App.~\ref{app:gamma}).

Although every analysis in this paper reads the spectrum averaged over the 126 segments, we keep the modulation in the simulations because real data contain it.
For this nearly four-year observation, averaging leaves the mean PSD almost
unchanged, while modulation increases its scatter where the Galaxy dominates. The simulations reproduce this effect
(Sec.~\ref{sec:gamma}), and correcting the likelihood for it changes the median posterior widths over
the 100 test simulations by less than 2\% (Sec.~\ref{sec:exact}, App.~\ref{app:gamma}).
Averaging discards the time pattern of the segments. With the depths unknown
this costs only 1--4\% in precision, although it would cost a factor of 2.1 in
$\sigma(\OGal)$ at one year if the depth were known exactly
(App.~\ref{app:gamma}). The averaged spectrum is therefore adequate here.
The network and SBI are trained on simulations with
the depths drawn at random, and the SBI posterior therefore allows for every
depth the prior permits, without the modulation being modelled by hand.

The extragalactic foreground, from inspiralling compact binaries, follows
\begin{equation}
  h^2\Omega_{\rm Ext}(f)=\OExt\,(f/1\,{\rm mHz})^{2/3}.
  \label{eq:ext}
\end{equation}
For the phase transition we use the broken power law of
Ref.~\cite{Caprini:2024} for bubble collisions and highly relativistic fluid
shells (its Template~I), with the spectral exponents of a strong transition,
\begin{equation}
  h^2\Omega_{\rm PT}(f)=\Ob\Big(\frac{f}{\fb}\Big)^{n_1}\Big[\frac12+\frac12\Big(\frac{f}{\fb}\Big)^{\aone}\Big]^{(n_2-n_1)/\aone},
  \qquad n_1=-n_2=2.4,\quad \aone=1.2.
  \label{eq:bpl}
\end{equation}
Its height $\Ob$, the value of the spectrum at its peak frequency $\fb$, and
$\fb$ itself relate to the transition strength $\alpha$, inverse duration
$\beta/H_*$ and temperature $T_*$ by
\begin{equation}
  \Ob=h^2F_{\rm GW,0}\,A_{\rm str}\,\tilde K^2\Big(\frac{H_*}{\beta}\Big)^2,\qquad
  \fb\simeq0.11\,H_{*,0}\,\frac{\beta}{H_*},\qquad \tilde K=\frac{\alpha}{1+\alpha},
  \label{eq:thermo}
\end{equation}
\begin{equation*}
  H_{*,0}=1.65\times10^{-5}\,{\rm Hz}\,\Big(\frac{g_*}{100}\Big)^{1/6}\frac{T_*}{100\,{\rm GeV}},\qquad
  h^2F_{\rm GW,0}=1.64\times10^{-5}\Big(\frac{100}{g_*}\Big)^{1/3},
\end{equation*}
where $H_*$ is the Hubble rate at the transition, $H_{*,0}$ its value
redshifted to today, $g_*=106.75$ the number of relativistic degrees of
freedom, $F_{\rm GW,0}$ the redshift factor of the energy-density fraction, and
$A_{\rm str}\simeq0.05$ a normalisation of the template~\cite{Caprini:2024}.
Ref.~\cite{Caprini:2024} derived the template for $\alpha\gg1$, corresponding
to $\tilde K$ close to 1, and keeps $\tilde K$ free to cover other scenarios, as we do.
We use this relation to draw the network's training set (Table~\ref{tab:priors})
and to express the posteriors in thermodynamic terms
(Sec.~\ref{sec:benchmark}). Both inference methods work directly in $\theta$.
The four signal parameters are
$\theta=(\log_{10}\OGal,\log_{10}\OExt,\log_{10}\Ob,\log_{10}\fb)$. The
benchmark figures use the injection point of Ref.~\cite{Caprini:2024},
$\log_{10}\OGal=-7.84$, $\log_{10}\OExt=-12.38$, $\log_{10}\Ob=-11.5$ and
$\fb=1$\,mHz.

\subsection{The averaged spectrum and its statistical law}\label{sec:gamma}

Each segment is Fourier transformed on a frequency grid of spacing
$10^{-6}$\,Hz from $3\times10^{-5}$ to $0.5$\,Hz, giving about $5\times10^5$
frequencies per channel. The PSD estimated from one segment, its periodogram,
scatters around the true PSD, and its average over the $\Nd=126$ segments is
the \emph{measured PSD} $\Dbar_I(f)$. For stationary Gaussian noise, each
periodogram follows an exponential distribution with mean $S_I(f)$,
independently across frequencies and segments. The average of $\Nd$ such values
therefore follows a Gamma distribution,
\begin{equation}
  \Dbar_I(f)\sim{\rm Gamma}\big(\text{shape }\Nd,\ \text{mean }S_I(f)\big),
  \label{eq:gamma}
\end{equation}
and the simulator draws $\Dbar_I$ directly from this law. The Galactic
modulation makes the segments differ. The simulator then draws a Gamma law with
the true mean and variance, whose shape parameter, the effective number of
segments $\Neff$, falls below $\Nd$ where the Galaxy dominates
(App.~\ref{app:gamma}).
Table~\ref{tab:setup} summarises the set-up and Table~\ref{tab:priors} the
priors. The analysis prior is uniform in $\theta$, whereas the separating
network was trained on a different prior, defined through the thermodynamic
parameters (Sec.~\ref{sec:sbiintro}).

In the stationary case this single Gamma draw is exactly equivalent to drawing
and averaging the 126 segment periodograms. With modulation it matches only
their mean and variance, and the effect of this approximation on the
posteriors has not been quantified. The comparison in App.~\ref{app:gamma}
measures the effect of the extra scatter on the likelihood, using the same
simulated data.

\begin{table}[tbp]
\centering
\caption{Simulation set-up, common to all data sets. $U[a,b]$ is uniform and
$\mathcal N(\mu,\sigma^2)$ Gaussian.}
\label{tab:setup}
\small
\begin{tabular}{p{2.9cm}p{11.7cm}}
\toprule
Quantity & Value \\
\midrule
Duration & 4\,yr, $\Nd=126$ segments of $\tau=10^{6}$\,s \\ %
Band & $3\times10^{-5}$--$0.5$\,Hz, $\Delta f=10^{-6}$\,Hz \\ %
Channels & A, E, T (T through $N_{\rm T}$ only in the likelihood) \\
Network input & one averaged spectrum, 512 bins per channel \\ %
Spectral exponents & $n_1=-n_2=2.4$, $\aone=1.2$, fixed (Eq.~\eqref{eq:bpl}) \\ %
Noise amplitudes & $\Aacc,\Poms\sim\mathcal N(\text{nominal},(0.05\,\text{nominal})^2)$ per simulation \\ %
Modulation & $d_1\sim U[0,0.2]$, $d_2\sim U[0.3,0.7]$, $\phi_{1,2}\sim U[0,2\pi]$, start time uniform over one year \\ %
\bottomrule
\end{tabular}
\end{table}

\begin{table}[tbp]
\centering
\begin{threeparttable}
\caption{Parameter draws for the separating network and for inference. The
network-training column gives the generating ranges before loudness scaling
and the acceptance checks described in App.~\ref{app:training}. A dash means
not drawn.}
\label{tab:priors}
\small
\begin{tabular}{p{3.6cm}p{5.0cm}p{3.6cm}}
\toprule
Parameter & Network training\tnote{a} & Inference\tnote{b} \\
\midrule
Components present & four simulation types\tnote{c} & all \\ %
$\log_{10}\tilde K$ & $U[-5,0]$\tnote{d} & --- \\ %
$\log_{10}(\beta/H_*)$ & $U[0,5]$\tnote{d} & --- \\
$\log_{10}(T_*/{\rm GeV})$ & $U[0,8]$\tnote{d} & --- \\
$\log_{10}\Ob$ & from Eq.~\eqref{eq:thermo}\tnote{d} & $U[-12.5,-10.5]$ \\ %
$\log_{10}(\fb/{\rm Hz})$ & from Eq.~\eqref{eq:thermo}\tnote{d} & $U[-4.0,-2.0]$ \\
$\log_{10}\OGal$ & $U[-8.5,-7.0]$ & $U[-8.5,-7.0]$ \\
$\log_{10}\OExt$ & $U[-13.0,-11.5]$ & $U[-13.0,-11.5]$ \\
Signal loudness $\mathcal A$\tnote{e} & $[0.01,100]$, validation at $\mathcal A=1$ & $\mathcal A=1$ \\ %
\bottomrule
\end{tabular}
\begin{tablenotes}\footnotesize
\item[a] 20\,000 training and 1\,024 validation simulations.
\item[b] SBI training, development and test sets (18\,000\,/\,2\,000\,/\,400) and all calibration-test simulations.
\item[c] Each draw selects one of four simulation types with probability $1/4$, namely noise only, noise and transition, noise and both foregrounds, and noise with all three components.
\item[d] Prior of Ref.~\cite{Caprini:2024}, mapped by Eq.~\eqref{eq:thermo}, redrawn until $\fb$ lies in the band.
\item[e] Multiplies the energy density of every signal component present, not the noise (App.~\ref{app:training}).
\end{tablenotes}
\end{threeparttable}
\end{table}

\section{Inference methods}\label{sec:analyses}

\subsection{The exact likelihood and MCMC}\label{sec:exact}

Because the measured PSD at each frequency follows the Gamma law of
Eq.~\eqref{eq:gamma}, independently across frequencies and channels, the
log-likelihood is the sum of the logarithms of the Gamma densities,
\begin{equation}
  \lnL(\vartheta)=\sum_{I}\sum_{f}\Big[(\Nd-1)\ln\Dbar_I-\Nd\ln S_I(f;\vartheta)
  -\frac{\Nd\,\Dbar_I}{S_I(f;\vartheta)}+\Nd\ln\Nd-\ln\Gamma(\Nd)\Big].
  \label{eq:lnL}
\end{equation}
This is equivalent to the Gaussian likelihood of the Fourier coefficients
themselves~\cite{Guan:2025}, because that likelihood depends on the data only
through the PSD averaged over segments. It is evaluated at every frequency of
the grid, without combining neighbouring frequencies into coarser bins, an approximation used in
Ref.~\cite{Caprini:2024}.
In this likelihood, T contributes through its noise only, because its response
to the GW background is strongly suppressed (Sec.~\ref{sec:channels}). Near
even multiples of $c/2L$, however, the noise of T vanishes faster than its
response, and the signal is no longer small compared with the noise. We exclude
narrow windows around these frequencies. Each window has a half-width of 1\%
of its central frequency, set by an assumed 1\% arm-length mismatch, and
together they remove 4.8\% of the T-channel frequencies.
At low frequency T is dominated by optical-metrology noise, whereas A and E are
dominated by acceleration noise, and T contains almost no signal. It therefore
measures $\Poms$ without confusion with the signal and narrows its uncertainty
by a factor of 1.2.
The likelihood uses the stationary shape $\Nd$ at every frequency, although the
modulation lowers it to $\Neff$ at frequencies where the Galactic foreground
contributes a large fraction of the total power (Sec.~\ref{sec:gamma}).
Using $\Neff$ instead changes the median posterior widths over the 100 test
simulations by less than 2\% (App.~\ref{app:gamma}).

Each evaluation of Eq.~\eqref{eq:lnL} sums about $1.5\times10^{6}$ terms across
the three channels. Repeated likelihood evaluations account for most of the
computational cost, and we therefore report cost as the number of
evaluations. We run the ensemble
MCMC sampler of Ref.~\cite{ForemanMackey:2013} with 48 walkers and 1\,500
steps, that is 72\,000 likelihood evaluations, whereas the converged reference
run for the benchmark (Sec.~\ref{sec:benchmark}) uses 288\,000, about 71
autocorrelation times, above the 50 recommended in
Ref.~\cite{ForemanMackey:2013}.

Overlapping signal components can produce several likelihood peaks. As a
check, we compare numerical searches started at the prior centre and at
the injected parameters. The latter are used only for this simulation check.
We also compare a local Gaussian approximation around a fitted peak, with
covariance given by the inverse likelihood curvature (the Laplace
approximation). The results of these checks are reported in
Sec.~\ref{sec:benchmark}. Locating competing peaks is therefore the first task
in which a learned method might help.

\subsection{The separating network and SBI}\label{sec:sbiintro}

\subsubsection{Separating noise and signal}
\label{sec:separatorio}

We reimplement the multi-scale, multi-headed autoencoder proposed by
Einsle et al.~\cite{Einsle:2025}, with the architecture also used in
Ref.~\cite{Einsle:2026}, and adapt it from ground-based simulations to LISA.
The input consists of the measured PSDs in A, E and T, each averaged over
126 segments, since retaining the individual segment spectra provides little
additional information (App.~\ref{app:gamma}). Each averaged PSD is
coarse-grained into 512 frequency bins using fixed weighted averages of
neighbouring frequencies (App.~\ref{app:training}). The network thus receives
$3\times512=1\,536$ values, expressed as normalised logarithms. The three
spectra share the same frequency grid and enter the network as separate
channels. One-dimensional convolutional filters slide along the frequency
axis, combining information from neighbouring bins and across the three
channels. The spectra are therefore processed together without concatenating
their frequency axes. This coarse-graining is used only for the network
input. The exact likelihood is evaluated on the original frequency grid
without coarse-graining.

An encoder extracts features at several frequency resolutions, and a noise
decoder predicts the instrument-noise spectrum. A second encoder reads this
prediction. The network subtracts its noise features from the total features
and adds a learned correction. A signal decoder then reconstructs the GW
spectrum, also using the predicted noise. For each channel it returns an
estimated noise spectrum $\nhat_I$ and an estimated GW spectrum $\shat_I$,
each with 512 bins. The separator takes one 512-bin measured spectrum from
each of the A, E and T channels, for a total of 1\,536 input values. For each
channel it estimates two 512-bin spectra, one for instrument noise and one for
the combined GW signal, yielding 3\,072 output values. For SBI, we retain the
GW estimates in A and E and the noise estimates in all three channels. These
five spectra contain 2\,560 values, which we compress to 32
principal-component scores.
The GW output contains the Galactic foreground,
the extragalactic foreground and the phase-transition background together.
It does not distinguish these three components. 

During training, each simulated measurement is paired with the corresponding
expected instrument-noise and combined GW spectra. The network is trained to
recover these mean spectra, rather than the random noise and signal
realisations in the measurement. The loss penalises differences between the
predicted and target spectra, favours a smooth GW spectrum, and encourages
consistency between the measured spectrum and the sum of the predicted noise
and signal. The signal-regression term is downweighted where the injected
signal is weak relative to the measurement uncertainty. Signal targets below
$10^{-54}$\,Hz$^{-1}$ are set to this floor. For these targets, predictions
at or below the floor receive no signal-regression penalty.
Appendix~\ref{app:training} gives the full training details.

At inference, the trained separator estimates the instrument-noise and
combined GW spectra from a new measured spectrum. Its intended domain is the
parameter region covered by the training simulations
(Table~\ref{tab:priors}), and its performance outside this region has not been
assessed.

The simulation settings and generating parameter ranges are given in
Tables~\ref{tab:setup} and~\ref{tab:priors}. We use 20\,000 training and
1\,024 validation simulations, mixing noise-only, transition-only,
foreground-only and combined GW injections, always including instrument
noise. A loudness curriculum progresses from loud to faint signals. Three
networks with different initial weights and data order are trained, and their
predictions are averaged in logarithmic power. For simulations that include a
phase-transition signal, we draw its thermodynamic parameters from the ranges
in Table~\ref{tab:priors} and map them to the signal amplitude and peak
frequency using Eq.~\eqref{eq:thermo}. This produces a broader range of
phase-transition signals than the analysis prior, which is uniform in the four
log signal parameters within the ranges in Table~\ref{tab:priors}. About 5\%
of the phase transitions used to train the separator fall within the analysis
ranges. This broad prior covers the range of transitions LISA might observe
and avoids tuning the separator to the analysis range. This choice
may reduce accuracy within the analysis range compared with training on that
range alone.

\subsubsection{From separated spectra to SBI}

For fixed physical parameters, the separator's predictions vary with the
random simulated measurement. Their distribution has no known analytical
form, unlike the Gamma law of the measured PSD. A simple analytical likelihood
for the network output is therefore unavailable.

SBI~\cite{Cranmer:2020} avoids this problem by learning the posterior from
simulations. It reads a summary statistic, a short set of numbers computed from
the data, and the more information this summary retains, the narrower the
posterior can be. For each SBI training simulation, we combine the five
retained spectra into a vector of 2\,560 values. We standardise these vectors
across the training set and use principal-component analysis (PCA) to identify
the 32 directions with the largest variance. The projections onto these
directions form the 32-value summary used by SBI. We exclude the
estimated GW spectrum in T from this summary because the GW power in T is
less than 1\% of the instrument-noise power in every frequency bin. %
We draw 18\,000 parameter sets from the analysis prior, simulate the data, pass
each simulated spectrum through the trained separator, and train a normalising
flow~\cite{Durkan:2019,Papamakarios:2016,Greenberg:2019}, a neural network that
represents a flexible probability distribution, to return the posterior of
$\theta$ given the summary statistic. The noise amplitudes and the modulation
are drawn in each simulation without being inferred, and the posterior is
therefore marginalised over them. The estimator is amortised. It is trained
once over the whole analysis prior and then returns the posterior for any new
data set at negligible cost, with no further training or likelihood evaluation.
Since SBI is trained on outputs from the fixed network using simulations drawn
from the analysis prior, its posterior reflects that prior regardless of how
the network was trained. Training the network on a broader prior may widen the
SBI posterior without biasing its uncertainty estimates.

We denote the estimator using the separated summary by \emph{SBI-S}. For
comparison, \emph{SBI-R} is trained separately on 32 principal components of
the measured A, E and T spectra, before neural separation. Both infer the
same four signal parameters and use the same 18\,000 training observations.
We fit the standardisation and PCA projection separately for the two
estimators, using their respective spectral inputs. Both use the same type of posterior network, with a longer
period without improvement allowed before training stops for SBI-R
(App.~\ref{app:sbi}).
\par

\begingroup
\subsubsection{Refining SBI for an individual observation}
\label{sec:sequential}

The estimators above are trained once for use on many observations. We
also consider truncated sequential neural posterior estimation
(TSNPE)~\cite{Deistler:2022}, a truncation strategy also used in
Ref.~\cite{Alvey:2023} and applied there to a ratio estimator, which refines the inference for one
observation by generating additional simulations where its posterior lies. We
apply it separately to the benchmark and to the three-peak observation of
Sec.~\ref{sec:benchmark}, in five steps. 1) Analyse the observation with SBI-S
and SBI-R to obtain two initial posteriors. 2) Take the region containing
almost all the probability of each posterior and keep their union within the
analysis bounds, so that either estimator can point to parameter values to
explore. ``Truncated'' refers to this restricted parameter range, and the
measured frequency range is unchanged. 3) Draw 4\,000 signal-parameter sets
uniformly from this region and simulate their spectra, with random
instrument-noise amplitudes and Galactic modulation as before. Both estimators
use these simulations, one after separation and the other directly from the
measured spectra. 4) Retrain both estimators from scratch on all the
additional simulations generated in earlier rounds, keeping the separating network and
the scaling and compression rules fixed and leaving out the original
18\,000 simulations. 5) Use the updated posteriors to set the region for the
next round. Four rounds give training sets of 4\,000, 8\,000, 12\,000 and
16\,000 simulations, and we call the final estimators \emph{TSNPE-S} and
\emph{TSNPE-R}. Neither is corrected with the exact likelihood. For each
observation, 200 further simulations from the final region check how often
the 68\% and 95\% intervals contain the true parameters. We also repeat the
final TSNPE-R training once, with a longer stopping patience, on the same
16\,000 simulations. These checks assess performance within the region
selected for that observation, not calibration across the full analysis
prior. The sequential study was performed after the main analyses.

\endgroup

\subsection{SBI-guided exact inference}\label{sec:guided}

SBI provides a rapid estimate of the foreground and phase-transition
parameters, although its posterior is broader than that obtained with the exact
likelihood and need not assign the correct relative probabilities to competing
solutions. We therefore use the SBI posterior to identify plausible regions
of parameter space, then refine the inference using the original averaged
A, E and T spectra and the likelihood of Sec.~\ref{sec:exact}.

The guided analysis searches these regions for distinct solutions and samples
around them. The samples are then reweighted to target the posterior defined
by the exact likelihood and the prior. This is importance sampling, and the
tail index $\khat$ of the weights flags results that may be unreliable
(App.~\ref{app:sbi}). This step recovers information lost
through the neural summary and determines the relative probabilities of
alternative combinations of the cosmological background, astrophysical
foregrounds and instrument noise. The method targets the same posterior as
the reference MCMC analysis. SBI guides where to evaluate the likelihood,
while the original data determine the final constraints.

Concentrating the likelihood evaluations in these regions reduces the
computational cost of recovering the reference posterior. Numerical checks
flag cases in which the sampled regions may not adequately represent it.
Appendix~\ref{app:sbi} describes the sampling, weighting and checks. We compare
the resulting posteriors and computational cost in Sec.~\ref{sec:benchmark},
and assess the reliability of their uncertainties in
Sec.~\ref{sec:reliability}.

In practice the three analyses form one workflow. SBI gives a calibrated
posterior at negligible cost once trained, and, when the exact likelihood is available, the
guided analysis turns it into a close approximation of the exact posterior
at a fraction of the cost of MCMC.

\section{Results}\label{sec:results}

\begingroup

We evaluate how well each method constrains the signal parameters and compare
the resulting posteriors, computational costs and calibration.

\subsection{Information preserved by the separator}\label{sec:role}

To assess how much information survives compression, we examine three
representations of the same simulated observation. These are the full
averaged spectra analysed with the exact likelihood, the 512-bin measured spectra compressed directly to 32 principal
components (the raw summary), and the separator output compressed to 32
principal components (the separated summary). The latter retains the GW
estimates in A and E and the noise estimates in A, E and T. The exact
likelihood uses the original frequency grid, without this compression.
Figure~\ref{fig:fisher} compares local Gaussian uncertainty estimates from a
Fisher analysis, with the same approximation to the analysis prior in all
three cases. The calculation and prior prescription are given in
App.~\ref{app:fisher}. The raw-summary comparison is a supplementary check.

For the benchmark of Sec.~\ref{sec:components}, the one-dimensional Fisher
uncertainties from the raw summary are 1.5--2.2 times larger than those from
the separated summary. The raw summary also shows stronger correlations
between $\OGal$, $\Ob$ and $\fb$. Its leading components mainly describe
instrument noise and the Galactic foreground, leaving less information
about the weaker transition. The full exact likelihood gives the smallest
uncertainties. Separation therefore helps retain parameter information in a
short summary, although some precision is lost relative to the full spectra.
The additional parameter-point comparison is given in
App.~\ref{app:fisher}. Two other summaries were not tested. Because the
likelihood is known here, compressing the spectra with its parameter
derivatives (score compression~\cite{Heavens:2000,Alsing:2018}) would give six
numbers, one for each parameter of the likelihood, with no local loss of
information, and subtracting the nominal noise spectrum before compression
could narrow the gap between the raw and separated summaries. Both use the
noise model directly, whereas the separator learns the noise from simulations.

\begin{figure}[tbp]
\centering
{\includegraphics[width=0.784\textwidth]{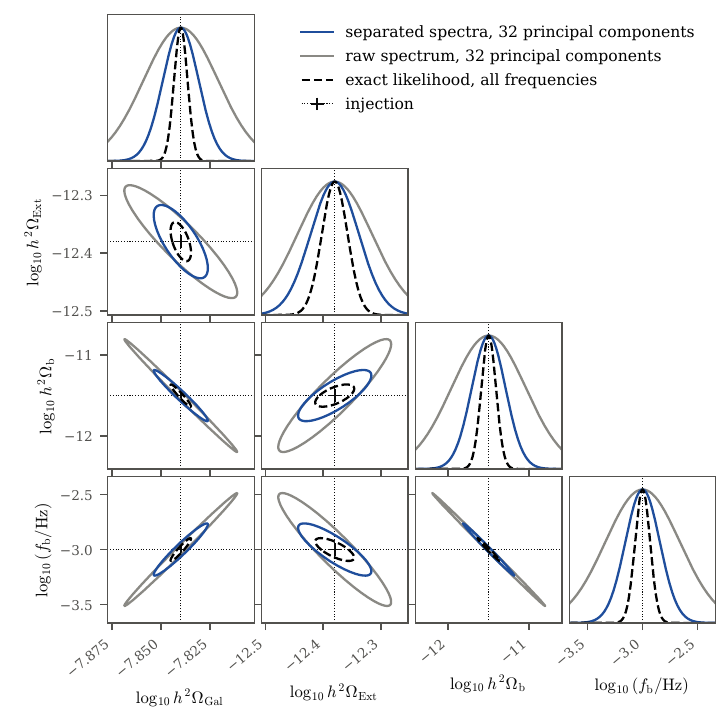}}
\caption{Fisher estimates for the benchmark observation. Blue contours use
the separated spectra compressed to 32 components, grey contours the measured
spectra compressed directly to 32 components, and black dashed contours the
full exact likelihood.
Off-diagonal contours enclose 68\% probability, and the diagonal panels show
one-dimensional Gaussian uncertainties. The noise amplitudes are
marginalised over. All three calculations include the same Gaussian
approximation to the uniform analysis prior, with matching variances. This
limits a weakly constrained direction in the raw summary. Crosshairs mark the
injected values. These curves are
local Gaussian estimates and not SBI posteriors. The prior prescription
and the comparison without it are given in App.~\ref{app:fisher}.}
\label{fig:fisher}
\end{figure}

We test whether the network's noise predictions improve inference by
comparing SBI trained on the GW outputs $\shat_{\rm A}$ and $\shat_{\rm E}$
alone with SBI trained on these outputs together with the noise outputs
$\nhat_{\rm A}$, $\nhat_{\rm E}$ and $\nhat_{\rm T}$. All five spectra are
predictions of the separating network (Sec.~\ref{sec:separatorio}). Here $\shat_I$
describes the combined Galactic foreground, extragalactic foreground and
phase-transition background, while $\nhat_I$ describes instrument noise.
Adding the noise outputs narrows the median posterior widths of the two
foreground amplitudes by about 20\%, of $\fb$ by 11\% and of $\Ob$ by 3\% on the
development set. The estimator without them, however, fails the calibration
test for $\OGal$, and the comparison is therefore only indicative. This supports including both the predicted GW and noise spectra
in SBI-S.

For spectral reconstruction, we compare the network prediction
$\shat_{\rm A}$ with a direct estimate of the GW power, obtained by
subtracting the nominal instrument-noise spectrum from the measured channel-A
spectrum. At each of 100 injected signal-parameter sets, labelled $k$, we
simulate 20 measurements, labelled $i$, and define
\begin{equation}
  \shat_{{\rm A},k}^{\rm sub}(f_j)
  =\max\!\left[\frac{1}{20}\sum_{i=1}^{20}\Dbar_{{\rm A},k,i}(f_j)
    -N_{\rm A}^{\rm nom}(f_j),\,s_{\min}\right],
  \label{eq:reconstruction-subtraction}
\end{equation}
where all spectra use the same frequency bins $f_j$, $N_{\rm A}^{\rm nom}$
is the noise model at its nominal amplitudes, and
$s_{\min}=10^{-54}$\,Hz$^{-1}$ is the floor applied before taking
logarithms. For the network, the three-network ensemble predictions
$\shat_{{\rm A},k,i}$ of Sec.~\ref{sec:separatorio} are averaged in
logarithmic power,
\begin{equation}
  \ell_{{\rm net},kj}=\frac{1}{20}\sum_{i=1}^{20}
      \log_{10}\shat_{{\rm A},k,i}(f_j),\qquad
  \ell_{{\rm sub},kj}=\log_{10}\shat_{{\rm A},k}^{\rm sub}(f_j),
  \label{eq:reconstruction-average}
\end{equation}
and both are compared with the injected combined GW spectrum
$s_{{\rm A},k}^{\rm true}(f_j)$,
\begin{equation} \epsilon_X=\operatorname{median}_{k,\,j\in\mathcal B}
      \left|\ell_{X,kj}-\log_{10}s_{{\rm A},k}^{\rm true}(f_j)\right|,
      \qquad X\in\{{\rm net},{\rm sub}\},
  \label{eq:reconstruction-error}
\end{equation}
where $\mathcal B$ contains the 148 frequency bins between 0.3 and
5\,mHz and $\epsilon_X$ is in dex, that is in decades of power. This gives $\epsilon_{\rm net}=0.081$\,dex
and $\epsilon_{\rm sub}=0.0039$\,dex, a ratio of 21. Averaging the logarithms
of the 20 individual subtraction estimates instead gives
$\epsilon_{\rm sub}=0.0040$\,dex and a ratio of 20.5. The ratio measures
spectral reconstruction error, not posterior interval width. The network
errors are smooth across frequency, and the Fisher comparison shows that the
compressed output still preserves useful information about the signal
parameters.

We also fit the separator output while assigning it the Gamma scatter of
the measured PSD, and compare the resulting Gaussian uncertainties and
interval coverage with the exact-likelihood reference. Depending on the signal
parameter, the median standard deviation across the 100 tests is between 2\%
and 49\% of the corresponding Laplace standard deviation from the
exact-likelihood analysis. The Laplace widths are computed at the highest peak
and therefore understate the width in the 42 simulations with several peaks.
The nominal 68\% intervals contain the injected value in only 0--1\% of the
tests.
The measured-PSD Gamma law therefore gives unreliable uncertainties for
the separated spectra, supporting the use of SBI to learn their posterior
from simulations.

\subsection{Recovered parameters and computational cost}\label{sec:benchmark}

\paragraph{Single-peak benchmark.}
We use one four-year simulated observation at the benchmark injection of
Sec.~\ref{sec:components}, $\theta=(-7.84,-12.38,-11.5,-3)$, with $\fb$
expressed in Hz, and one draw of the noise amplitudes and Galactic
modulation. Its exact posterior has one dominant peak.
Figure~\ref{fig:fig3analogue} compares the MCMC and SBI-guided posteriors.
MCMC uses 288\,000 likelihood evaluations and spans 71 autocorrelation times,
exceeding the convergence criterion in Sec.~\ref{sec:exact}. The relative uncertainties in
$\Aacc$ and $\Poms$ are 0.19\% and 0.005\%, respectively, and their
correlations with the signal parameters are weak ($|r|\le0.2$). The strongest
correlations are between signal components. The logarithmic transition
amplitude and peak frequency have $r=-0.98$, and the logarithmic Galactic and
extragalactic amplitudes have $r=-0.62$. The transition peak lies below the
Galactic foreground, and the data therefore mainly constrain its falling side. A lower
transition amplitude at a higher peak frequency can produce a similar
spectrum in this region.

SBI-guided inference recovers the same main correlations and similarly tight
noise constraints using 5\,130 likelihood evaluations. Its central 68\%
signal-parameter intervals are 10--15\% narrower than the MCMC intervals.
Both methods target the same posterior, and this difference must therefore be assessed
as a numerical discrepancy rather than a gain in physical information.
For this benchmark observation, the guided analysis draws
2\,000 candidate sets of the four signal and two instrument-noise parameters
from a distribution constructed around the solutions located using SBI.
These samples are reweighted using the exact likelihood and prior to estimate
the posterior (App.~\ref{app:sbi}). The number 2\,000 is the standard sampling
budget used for all observations, including the blind test of Sec.~\ref{sec:reliability}. Repeating the analysis of the same data with a
tenfold larger budget of 20\,000 parameter samples brings the interval widths
within 2\% of MCMC, at a total cost of about 23\,000 likelihood evaluations,
including the searches for solutions, although its $\hat k = 0.86$ then exceeds
the threshold of 0.7, flagging the heavier tail that the smaller sample did not
reach. The $\hat k$ check does not flag the underestimated widths in the
smaller run ($\hat k = 0.66$), showing that passing this
check does not guarantee accurate uncertainties for an individual observation.
The smaller sample also produces the less smooth guided contours in
Fig.~\ref{fig:fig3analogue}. The
inset expresses the benchmark constraints in terms of $\alpha$ and
$\beta/H_*$ at fixed $T_*=200$\,GeV, and the injected values are $\alpha=1.18$
and $\beta/H_*=272$.

\begin{figure}[tbp]

\centering
{\includegraphics[width=0.85\textwidth]{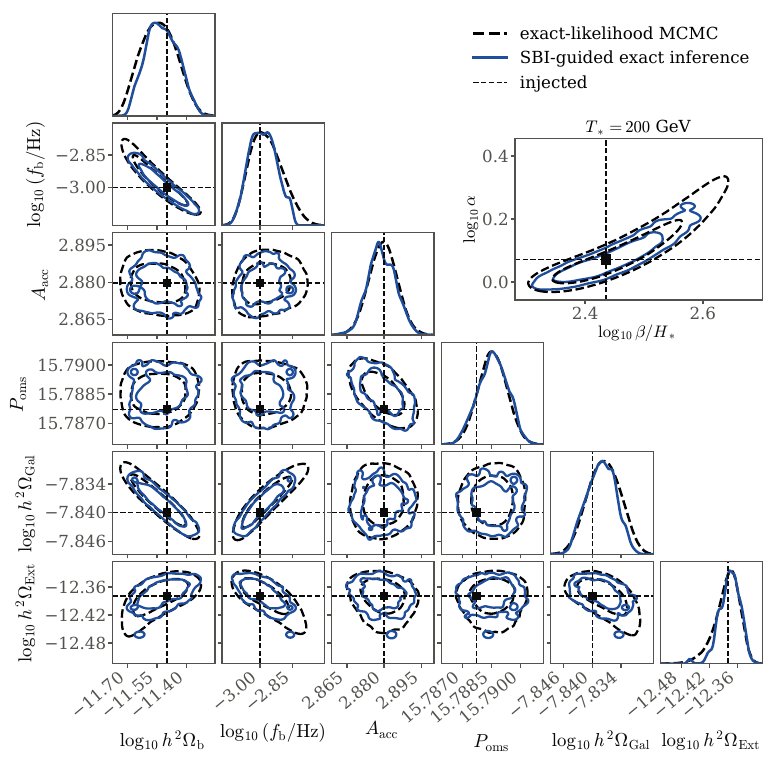}}
\caption{Benchmark posterior for the two instrument-noise amplitudes and four
signal parameters. Black dashed contours show exact-likelihood MCMC with
288\,000 likelihood evaluations (71 autocorrelation times), and blue contours
SBI-guided inference with 5\,130 likelihood evaluations, 2\,000 weighted draws and $\khat=0.66$,
with an effective sample size of about 590.
Contours enclose 68\% and 95\% probability. The black square and dashed lines
mark the injected values. The inset maps the signal posterior to $\alpha$
and $\beta/H_*$ at $T_*=200$\,GeV using Eq.~\eqref{eq:thermo}. This mapping
extrapolates the strong-transition template at the injection and elsewhere
outside $\alpha\gg1$~\cite{Caprini:2024}.}
\label{fig:fig3analogue}
\end{figure}

\paragraph{SBI-S and sequential refinement.}
We now consider the posterior returned directly by SBI-S, before the
exact-likelihood analysis used to obtain the guided result above. SBI-S is
trained once over the analysis prior and estimates the four signal
parameters from the separated 32-component summary
(Sec.~\ref{sec:sbiintro}). At the single-peak benchmark, its central 68\%
intervals are 3.9--6.2 times wider than the MCMC intervals, depending on the
parameter. We also compare SBI-S with the guided reference for each of the
100 blind-test observations. For each signal parameter, we calculate the
ratio of their interval widths in each observation and take the median
over the 100 observations. The four median ratios range from 3.1 to 5.6.
To assess whether additional training can reduce these broader intervals,
we next apply the sequential procedure, which concentrates new simulations
on the parameter regions relevant to an individual observation without
using an exact-likelihood correction.

The sequential procedure of Sec.~\ref{sec:sequential} improves the benchmark
constraints. Relative to the same MCMC intervals, the width ratios are
1.4--2.8 for TSNPE-S and 3.0--6.7 for TSNPE-R after its longer training run.
The TSNPE-S widths lie below the prior-free Fisher widths of the separated
summary at the benchmark (Table~\ref{tab:fisherpoints}) because the sequential
estimator includes the bounded analysis prior, and with that prior included
they are comparable to the summary's Fisher widths (Fig.~\ref{fig:fisher}).
Both pass the local coverage check on 200 further simulations, with TSNPE-R
passing after the additional training. This single retraining was fixed
before it was run, after the first TSNPE-R run had failed the local check by
one count. These local results
do not establish calibration across the full analysis prior.

\paragraph{Three competing solutions.}
The single-peak benchmark tests how precisely each method constrains one
solution. To test whether the methods can recover several plausible
solutions, we consider a separate simulated observation drawn from the
analysis prior (Table~\ref{tab:priors}). For this observation, different
combinations of foreground amplitudes and phase-transition parameters
produce similar total spectra, giving three distinct regions of high
posterior probability in signal-parameter space. In Fig.~\ref{fig:corner},
SBI-guided inference locates the same three regions as MCMC, while the
broader SBI-S posterior covers all three. To compare the relative support
for these solutions across MCMC, guided inference, SBI-S and the sequential
estimators, we need the probability assigned to each region under a common
definition. We define the three regions by fitting three Gaussian
components to the MCMC samples in the four signal parameters. Samples from
every method are then assigned to the closest component, accounting for its
widths and parameter correlations when measuring distance. Using this common
assignment, Table~\ref{tab:regions} compares the posterior probability in each
region, including the importance weights for guided inference. Guided
inference assigns 52.8, 34.9 and 12.2\% to the three regions, compared with
53.2, 35.3 and 11.5\% for MCMC, consistent within the quoted sampling errors.
In contrast, SBI-S, TSNPE-S and TSNPE-R assign the third region about 3--5
times the MCMC probability. The first two regions lie along the same ridge
relating the phase-transition amplitude to its peak frequency, and hence their
boundary depends on the assignment rule. Their individual probabilities
should therefore be interpreted as a comparison under this fixed partition. The marginal interval widths reported
above measure precision in individual parameters.

\begin{figure}[tbp]
\centering
{\includegraphics[width=0.7\textwidth]{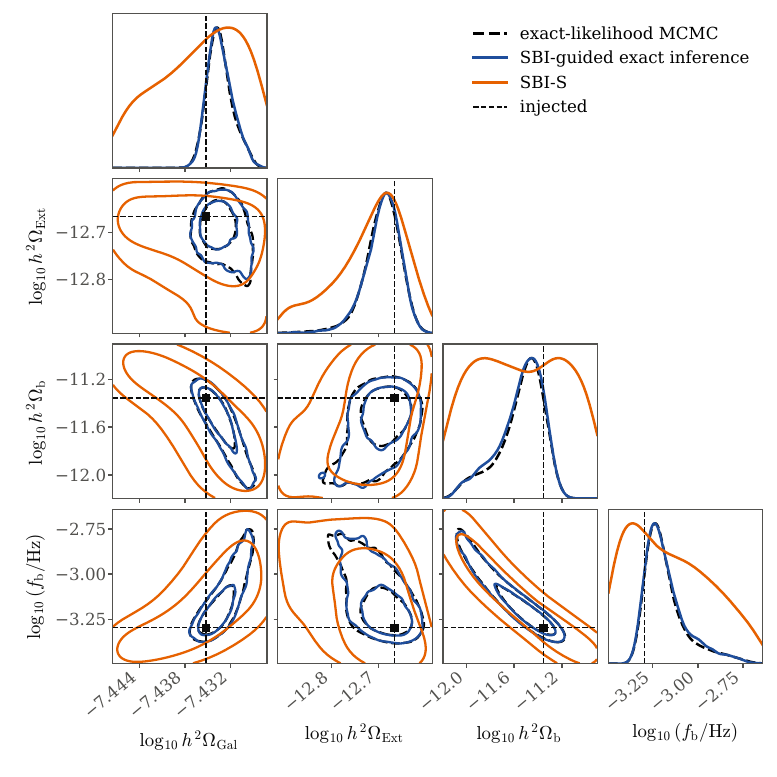}}
\caption{The separate test observation with three posterior solutions.
Black dashed contours show exact-likelihood MCMC with 180\,240 retained
samples, blue contours SBI-guided inference and orange contours SBI-S.
Contours enclose 68\% and 95\% probability. The guided posterior resolves the three
regions found by MCMC. The broader SBI-S posterior assigns probability to
all three regions. These are alternative parameter combinations describing
the same observation, rather than three peaks in the frequency spectrum.}
\label{fig:corner}
\end{figure}

The search tests in Sec.~\ref{sec:exact} also show the importance of checking
for alternative solutions. The searches started at the prior centre and at
the injected parameters end at different likelihood peaks in 42 of 100
simulations.
A Gaussian approximation around one fitted peak can underestimate the
uncertainty by up to a factor of two. Among these 42 cases, starting at the
SBI posterior mean reaches the best known peak in 29 cases, compared with
19 when starting at the prior centre. We did not compare the guided search
with a multistart search that does not use SBI.

\paragraph{Likelihood cost.}
For the same 100 blind-test observations used in the interval-width
comparison, the final SBI-guided procedure uses a median of 4\,920 exact
likelihood evaluations per observation, with a 10--90\% range of
3\,680--6\,046 (Table~\ref{tab:compare}). This count includes the peak
searches, local curvature estimates and importance weighting. The median
is about one fifteenth of the nominal MCMC budget of 72\,000 evaluations
and about one sixtieth of the 288\,000 evaluations of the converged
benchmark run (Sec.~\ref{sec:exact}). The comparison excludes simulation and
network training costs. These results use SBI-S as the guide. The
calibration of SBI-R is discussed in Sec.~\ref{sec:reliability}.

\begin{table}[tbp]
\centering
\caption{Summary of the three analyses. SBI-S width ratios are parameter-wise
medians of central 68\% interval-width ratios over the 100 guided-test
observations, which are a subset of its 400 test observations. SBI-S
calibration uses all 400. The guided width ratios refer only to the single-peak benchmark and use
MCMC as the reference. Its likelihood cost is the median and 10--90\% range
over the 100 blind-test observations. Simulation and training costs are
excluded.}
\label{tab:compare}
\small
\begin{tabular}{L{3.2cm}L{3.6cm}L{3.2cm}L{3.7cm}}
\toprule
Analysis & Calibration result & 68\% interval-width comparison & Likelihood evaluations \\
\midrule
Exact-likelihood MCMC & reference & 1 (reference) & 72\,000 nominal, 288\,000 converged (benchmark) \\
SBI-S & passes on 400 test observations & 3.1--5.6 times guided reference (100 observations) & none after training on 18\,000 simulations \\
SBI-guided exact inference & passes 12 checks on 100 blind-test observations & 0.85--0.90 times MCMC (benchmark only) & 4\,920 (3\,680--6\,046) \\
\bottomrule
\end{tabular}
\end{table}

\subsection{Reliability of the reported uncertainties}\label{sec:reliability}

We assess calibration by checking how often the inferred intervals contain
the injected parameters. A central 68\% interval should contain the true
value in approximately 68\% of observations, and similarly for a 95\%
interval, allowing for finite-sample fluctuations. Test observations have
signal parameters drawn from the analysis prior, with random noise
amplitudes and Galactic modulation. The injected values are not supplied
to the inference. SBI is tested on 400 observations, and the final guided
procedure on 100 new observations not used in its development. Acceptance
criteria are fixed before the corresponding test samples are generated.
The tests at other probability levels and jointly across the four signal
parameters, together with their thresholds, are defined in
App.~\ref{app:sbi}.

On the 400 test observations, SBI-S passes the 68\% and 95\% interval-coverage
checks, the parameter-by-parameter simulation-based calibration (SBC) tests,
and the joint four-parameter test of accuracy with random points
(TARP)~\cite{Lemos:2023}. Figure~\ref{fig:pp} (left) shows the interval-coverage curves, and
Table~\ref{tab:calib} gives the numerical test results. On the larger development
set of 2\,000 observations, however, a small calibration error is detected
in one parameter, and the affected parameter changes when training starts from
different random weights. SBI-S therefore passes the stated test criteria,
although the larger sample reveals residual calibration errors.

SBI-R fails the parameter-by-parameter calibration test for $\OGal$, including
after longer training. Its interval widths are therefore excluded from the
comparison of estimators trained once over the full prior. This result is
separate from the local TSNPE-R check in Sec.~\ref{sec:sequential}, which uses
new training simulations concentrated around one observation.

\begin{figure}[tbp]
\centering
{\includegraphics[width=0.98\textwidth]{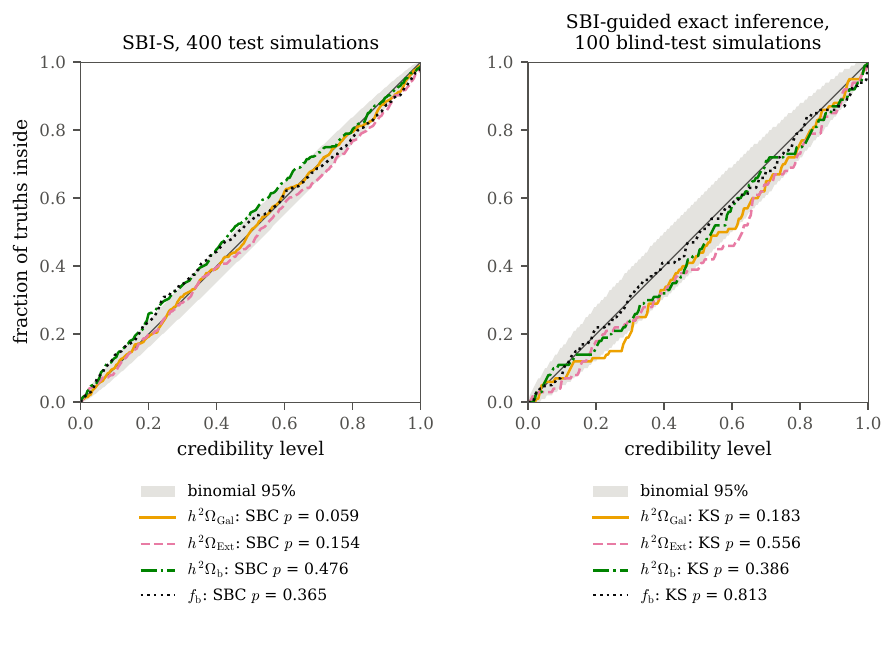}}
\caption{Interval coverage for each signal parameter. The horizontal axis is
the stated probability of a central credible interval, and the vertical axis
is the fraction of observations whose injected value lies in that interval.
A calibrated method follows the diagonal, and the shaded band shows the 95\%
range expected from finite-sample binomial scatter. The left panel shows
SBI-S on 400 test observations, and the right panel SBI-guided inference on
100 blind-test observations. Legends give the pre-specified calibration-test
$p$-values, namely SBC for SBI-S
and the Kolmogorov--Smirnov test for guided inference. The thresholds are
given in App.~\ref{app:sbi}.}
\label{fig:pp}
\end{figure}

The final SBI-guided procedure passes all twelve pre-specified checks on
100 new blind-test observations (Fig.~\ref{fig:pp}, right, and
Table~\ref{tab:calib}). There are three checks for each of the four signal parameters, namely 68\%
coverage, 95\% coverage and the calibration test across probability levels, a
Kolmogorov--Smirnov test of the probability integral transform (PIT), the
posterior probability below the injected value. All eight coverage counts lie
at or below their nominal values, consistent with the slightly narrow intervals
found for the benchmark at 2\,000 draws (Sec.~\ref{sec:benchmark}). The
main test includes all 100 outputs, including 18 flagged as unreliable by
the importance-weight check. The remaining 82 outputs also pass when
tested separately. Passing these ensemble checks does not establish the
accuracy of each flagged posterior. The final procedure was fixed after
development on earlier simulations, as documented in App.~\ref{app:sbi}.

\endgroup
\section{Discussion and conclusions}\label{sec:conclusions}

We explored a machine-learning approach to inferring the parameters of stochastic GW
backgrounds in simulated LISA data in which a separating network splits each measured spectrum into instrument noise and the
combined GW signal, and SBI turns a compressed form of this output into a
posterior. Exact-likelihood MCMC on the same simulated data served as the
reference. In a Fisher analysis without the prior, 32 principal components of
the separated spectra give signal-parameter uncertainties comparable to those
from about 300 principal components of the raw spectra, and the separated
summary's widths are 1.6--2.2 times the exact-likelihood widths, taking the median for each parameter over ten selected
cases (Sec.~\ref{sec:role}, App.~\ref{app:fisher}).
The amortised SBI estimator returns a calibrated posterior at negligible cost
once trained, with uncertainties three to six times wider than the exact ones.
The compression and the estimator contribute comparably to this extra width,
and the sequential estimator removes most of the estimator's share at the
benchmark (Sec.~\ref{sec:benchmark}).
Learning the uncertainties is essential, because treating the network output
as data with the scatter of the measured PSD gives intervals that almost never
contain the true values (Sec.~\ref{sec:role}).
Used as a guide to the exact likelihood, the SBI posterior locates competing
peaks, flags cases where one may have been missed, and closely
approximates the exact posterior with 15--60 times fewer likelihood
evaluations than MCMC
(Secs.~\ref{sec:benchmark} and~\ref{sec:reliability}).
A separating network trained on the analysis prior itself could reduce the
cost of the separation, although it would not address the larger loss in the
estimator.

\subsection{Degeneracies in phase-transition inference}\label{sec:noise}

In these simulations, the main difficulty is distinguishing
the phase-transition background from the astrophysical foregrounds.
Instrument noise is already tightly constrained by the exact-likelihood
analysis, with the T channel providing a largely independent measurement of
noise. At the benchmark, both noise amplitudes are determined to better than
one per cent and are only weakly correlated with the signal parameters
(Sec.~\ref{sec:benchmark}). Different combinations of foreground amplitudes
and phase-transition parameters can nevertheless produce similar total GW
spectra. This ambiguity leads to correlated parameter uncertainties and, in
some observations, competing posterior solutions. Separating instrument noise
from the combined GW signal does not distinguish these overlapping GW
components.

The separating network also uses the same measured spectra as
the exact-likelihood analysis. Its output can preserve or discard information
from those spectra. It cannot, however, add information. Within the assumed data
model, noise separation therefore offers no additional information with which
to improve the phase-transition constraints beyond those of the reference
analysis. Its demonstrated benefits are compact summaries for SBI and guidance
for faster likelihood-based inference.

\subsection{Outlook and limitations}\label{sec:outlook}

These results hold for one family of simulators. They assume equal-arm
first-generation TDI, which makes the noise of the three channels uncorrelated.
They also assume stationary Gaussian instrument noise and a Galactic
modulation described by one common factor $m(t)$ at every frequency and in
every channel. This simplifies the channel-dependent modulation discussed in
Sec.~\ref{sec:components}. The exact
likelihood uses T for its noise only, and its stationary form is a few per cent
overconfident in the Galactic amplitude when the foreground is modulated
(App.~\ref{app:gamma}). The guided procedure was tested blind on 100 simulations,
which cannot exclude a width error of about 10\% in one parameter.
Whether the separated summary retains its advantage over the raw summary
under non-stationary noise remains to be tested. Likelihood-based score
compression~\cite{Heavens:2000,Alsing:2018} provides another useful comparison,
although it requires an explicit likelihood model.

Where the likelihood is correct, learned methods make exact inference faster
and more robust rather than more precise. They may outperform a hand-built
likelihood where it is wrong or cannot be written, for a Galactic foreground
whose statistics follow from a population of binaries, for non-stationary
noise, glitches and data gaps~\cite{Baghi:2022}, or for a transition spectrum
that is not a broken power law. The time sequence of segment spectra would also
become informative if the noise varied between segments or the modulation
depth were known independently~\cite{Criswell:2024}. The
analyses here provide a controlled baseline, with uncertainties calibrated
under criteria fixed in advance and a known distance from the best possible
precision, against which such methods can be tested.

\section*{Data availability}

The code for generating simulated data, training the networks, and running
inference is available on GitHub~\href{https://github.com/AI-Driven-HEP/lisa-gwb-inference}{\faGithub}.

\appendix

\section{The Gamma law and the Galactic modulation}\label{app:gamma}

Without Galactic modulation, averaging $\Nd$ independent exponential
periodograms gives the Gamma law of Eq.~\eqref{eq:gamma} exactly. With
modulation, the segment means
$S_I^{(j)}=S_{I,{\rm stat}}+m_jS_{I,{\rm Gal}}$ differ. Here $S_{I,{\rm Gal}}$
is the Galactic contribution to $S_I$ before modulation, $S_{I,{\rm stat}}$ the
remaining stationary part, and $m_j$ the modulation of
Eq.~\eqref{eq:modulation} averaged over segment $j$. The exact distribution of
their average is generally not Gamma. The simulator instead uses a Gamma
approximation with the same mean and variance~\cite{Welch:1947,Satterthwaite:1946}.
\begin{equation}
  \begin{aligned}
  Q_1&=\sum_{j=1}^{\Nd}S_I^{(j)},\qquad
  Q_2=\sum_{j=1}^{\Nd}\big(S_I^{(j)}\big)^2,\\
  \mathbb E[\Dbar_I]&=\frac{Q_1}{\Nd},\qquad
  \operatorname{Var}(\Dbar_I)=\frac{Q_2}{\Nd^2},\qquad
  \Neff=\frac{Q_1^2}{Q_2}.
  \end{aligned}
  \label{eq:neff}
\end{equation}
Thus $\Neff$ is the Gamma shape required to reproduce these two
moments. Tests using $10^6$ simulated values recover the mean and variance
within Monte Carlo uncertainty. Higher moments need not agree. At a purely
Galactic frequency and the deepest modulation allowed by the prior
($\Neff=100$), the Gamma skewness is 9.8\% below that of the exact average.

The reference likelihood of Eq.~\eqref{eq:lnL} uses the stationary shape
$\Nd$. Replacing it by $\Neff(f,\vartheta)$ accounts for the simulator's
modulation-dependent Gamma scatter. This correction retains the reference
analysis's neglect of the weak GW contribution in T. It does not replace the
Gamma approximation by the exact distribution of the segment average. From the fiducial depth to the deepest modulation
of the prior, a Fisher calculation that includes the larger scatter shows the
stationary likelihood understating the standard deviation by 2.4--4.8\% for
$\log_{10}\OGal$ and by 1.7--3.5\% for the transition parameters. In a simpler
test on 400 deeply modulated simulations, varying only the Galactic amplitude,
the scatter of the fitted Galactic amplitude across simulations is
$1.107\pm0.035$ times its estimated uncertainty, compared with
$1.024\pm0.035$ after the modulation correction. Over the 100 test simulations,
switching the SBI-guided posteriors to the corrected likelihood changes the
median width of $\log_{10}\OGal$ by 1.6\% and the others by at most 0.2\%, and
no calibration count by more than four. We therefore use the stationary form.

For a stationary Gaussian data model, the averaged PSD contains all the
segment spectra's information about their common mean PSD. A control network
using the average performed as well as one using the full sequence.
Modulation can make the time sequence informative, although its varying part
constrains the products of the Galactic amplitude and the modulation depths.
When the depths are unknown, this limits how much timing information helps
separate the Galactic foreground from the other GW components. Under our
prior, a Fisher comparison of the 126 segment spectra with their average
finds only a 1--4\% improvement in parameter uncertainties from the full
sequence. In a separate one-year comparison with the depths known exactly,
the Galactic-amplitude uncertainty would improve by a factor of 2.1. Most of
that gain requires the depths to be known to about one per cent.

\section{The separating network and its training}\label{app:training}

Section~\ref{sec:separatorio} describes the architecture and training targets.
Here we specify the binning, loss and curriculum used to train the separator,
and Table~\ref{tab:hyper} lists the numerical settings.

Each channel's measured PSD is first coarse-grained onto 512 bins, each a
weighted average of the frequencies it contains. In A and E the weight of a
frequency is $(R_I/f^3N_I^{\rm fid})^2$, where $N_I^{\rm fid}$ is the noise at
nominal amplitudes. This weight is proportional to the local Fisher information about a signal
amplitude in the noise-dominated limit, and to $1/\Omega_{\rm sens}^2$, with
$\Omega_{\rm sens}$ the sensitivity of Fig.~\ref{fig:sensitivity}. The
frequencies where LISA is most sensitive therefore dominate their bin. In T,
which carries almost no signal, the weight is
$(\partial\ln N_{\rm T}/\partial\ln\Aacc)^2+(\partial\ln N_{\rm T}/\partial\ln\Poms)^2$,
the information the frequency carries about the two noise amplitudes.
We compute the binning weights once from the reference instrument model and
apply the same weights to every simulated spectrum. In A and E we use the noise
spectra at the nominal amplitudes, and in T we evaluate the noise derivatives
at those same amplitudes.

The network uses the dimensionless normalisation
$(\log_{10}X+45)/10$, where $X$ is the numerical PSD in Hz$^{-1}$.
One normalised unit therefore corresponds to ten decades of power. The signal
floor and its one-sided regression penalty are defined in
Sec.~\ref{sec:separatorio}. The floor, $10^{-54}$\,Hz$^{-1}$, is ten times
below a level exceeded by the peak signal in 95\% of simulations at physical
loudness ($\mathcal A=1$, defined below). It is a training convention rather
than an observation-specific detection threshold.
The total loss is the weighted sum of spectral, latent-feature and
recombination terms, with coefficients given in Table~\ref{tab:hyper}.
The spectral term includes a detectability weight for signal regression,
$w=\rho^2/(1+\rho^2)$, where $\rho=[S/(S+N)]/\sigma_{\rm frac}$ and
$\sigma_{\rm frac}$ is the fractional uncertainty of a coarse-grained
measured bin. Here $S$ and $N$ are the simulated signal and noise targets.
This reduces the demand to reconstruct signals that the input cannot measure.
This weight applies only to signal regression above the floor. The noise
error and the one-sided penalty at the floor are unchanged.
The additional losses compare internal features of predicted and true noise,
and features of the recombined prediction and measured input. The
recombination is performed in physical power before returning to logarithms.
These are training conditions, not an exact subtraction imposed on the output.
The training-pool builder also rejects and redraws examples whose normalised
noise or signal targets fall outside $[-1,1]$.

\begin{figure}[tbp]
\centering
\includegraphics[width=0.48\textwidth]{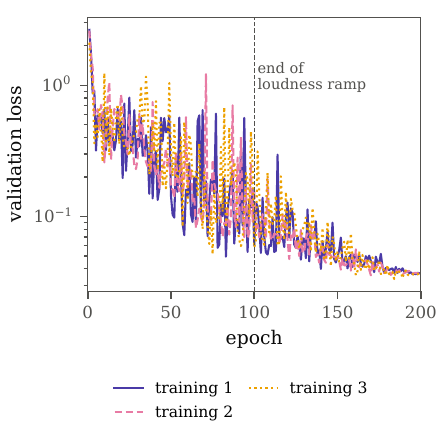}
\caption{Validation loss of the three trainings at $\mathcal A=1$. No training
loss is shown, because the training objective is curriculum-weighted, and its value is therefore
not comparable from epoch to epoch or with the validation loss.}
\label{fig:loss}
\end{figure}

Following Ref.~\cite{Einsle:2025}, the network learns first from loud signals
and then from fainter ones. A loudness factor $\mathcal A$ multiplies the energy
density of every signal component and leaves the noise unchanged. At epoch $e$,
$\mathcal A$ is drawn from a log-normal distribution $p_e(\mathcal A)$, clipped
to $[\mathcal A_{\min}(e),100]$, with $\mathcal A_{\min}$ falling log-linearly
from 100 to 0.01 over the first 100 of the 200 epochs. In the last 100 epochs
the median $\mathcal A$ is 0.05, and the network therefore finishes training on
signals mostly fainter than the physical ones ($\mathcal A=1$). Rather than
simulate new data at every epoch, we draw the 20\,000 training simulations once
from the average of these distributions over all epochs, $\bar
p(\mathcal A)$. At epoch $e$, simulation $i$ receives the weight
$w_i=p_e(\mathcal A_i)/\bar p(\mathcal A_i)$, which makes the weighted training
set follow $p_e(\mathcal A)$. No weight can exceed the number of epochs, and the
largest is 78. The network is validated on 1\,024 separate simulations at
$\mathcal A=1$, without weights (Fig.~\ref{fig:loss}).

A scan of five learning rates showed a flat range around the rate used. At epoch 200, the three runs have validation losses within 1.7\% of one
another (Table~\ref{tab:hyper}). Their signal
output grows in proportion to the injected signal, with slopes of 0.94--0.96
over the frequencies where the signal is detectable, and at no frequency is the
output stuck at a constant value.

\begin{table}[tbp]
\centering
\caption{The separating network and its three production trainings, which
differ only in random initialisation and data order. Loss is the spectral part
of the validation loss. The best loss over epochs is biased low, and the loss at
epoch 200 is therefore the one compared. The weights $\lambda_s$, $\lambda_l$
and $\lambda_c$ multiply the error of the output spectra, a term comparing the
internal features of the predicted and true noise, and a term comparing the
features of their recombination with those of the measured spectrum. The smoothness
weight penalises curvature in the signal output.}
\label{tab:hyper}
\small
\begin{tabular}{ll}
\toprule
Encoder levels & 3 \\ %
Channels at first level & 64 \\
Trainable parameters & 1\,672\,518 \\ %
Input & One averaged spectrum, A, E and T, 512 bins each \\ %
Loss weights $(\lambda_s,\lambda_l,\lambda_c)$ & $(0.5,0.1,0.1)$ \\ %
Smoothness weight & 1.0 \\
Optimiser & Adam, cosine schedule \\
Learning rate & $3.6\times10^{-4}$ \\ %
Weight decay & $5\times10^{-6}$ \\
Batch size & 64 \\ %
Epochs & 200 \\
Gradient clipping & unit norm \\
Training / validation simulations & 20\,000 (curriculum) / 1\,024 ($\mathcal A=1$) \\ %
\midrule
\multicolumn{2}{l}{\begin{tabular}{@{}lcccc@{}}
Training & best loss & epoch of best & loss at epoch 200 & slope \\
1 & 0.0364 & 195 & 0.0372 & 0.949 \\
2 & 0.0357 & 170 & 0.0372 & 0.944 \\
3 & 0.0335 & 184 & 0.0366 & 0.955 \\
\end{tabular}} \\
\bottomrule
\end{tabular}
\end{table}

\section{SBI and the SBI-guided procedure}\label{app:sbi}

\subsection{Posterior estimation and sequential refinement}

SBI uses neural posterior estimation~\cite{Papamakarios:2016,Greenberg:2019}. A
network $q_\psi(\theta|s)$, with weights $\psi$, represents a probability
distribution of the signal parameters $\theta$ that depends on the summary
statistic $s$ of the data. It is trained on pairs of parameters $\theta$ drawn
from the analysis prior $\pi$ and data $\mathbf d$ simulated from them, by minimising
\begin{equation}
  J(\psi)=-\mathbb E_{\theta\sim\pi,\,\mathbf d\sim p(\mathbf d|\theta)}\big[\ln q_\psi(\theta|s(\mathbf d))\big].
  \label{eq:npe}
\end{equation}
The loss $J$ is the average negative log density that the network assigns to
the injected parameters, given their simulated data summaries. If the density estimator were unrestricted and trained on
infinitely many simulations, the minimum would recover
$p(\theta|s(\mathbf d))$, the posterior conditional on the chosen summary. The finite
network and training set approximate this limit.

SBI-S uses the five separated spectra specified in
Sec.~\ref{sec:sbiintro}. SBI-R uses the three measured spectra. Each input is
standardised and projected onto its own 32 leading principal components, each
scaled to unit variance.
A small fully connected embedding network then supplies the input to
$q_\psi$. It is trained jointly with a neural spline
flow~\cite{Durkan:2019} using the \texttt{sbi}
package~\cite{TejeroCantero:2020}. A transformation from the bounded signal-prior
box to the real line keeps the inferred parameters inside the prior bounds.
The amortised estimators use 18\,000 simulations for fitting, including their
internal validation split, and separate sets of 2\,000 development and 400
test simulations.

The two inputs contain 2\,560 separated or 1\,536 measured values before
compression. Standardisation and PCA are fitted on the 18\,000-simulation
training set before the internal validation split, without using the separate
development or test sets. Both estimators use the same flow and embedding
architecture, learning rate ($5\times10^{-4}$), batch size (256), and internal
validation fraction (0.1).
Training stops after the validation performance has failed to improve for
20 consecutive epochs for SBI-S and 60 for SBI-R. An epoch is one pass through
the training set.

In each TSNPE round, the retained region is the union of the regions containing
99.99\% of the probability of the two current posteriors, restricted to the
analysis bounds. The flow and the small embedding network after PCA are
retrained from scratch, while the separator, standardisation and PCA
projections remain fixed. Both estimators use the SBI-S settings, including
the stopping rule of 20 epochs without improvement. The additional TSNPE-R
run uses 60 epochs without improvement on the same final 16\,000 simulations.
Section~\ref{sec:sequential} specifies the simulation budget and local
coverage checks for this observation-specific procedure.

\subsection{Guided likelihood sampling}\label{app:guided}

SBI-guided inference uses importance sampling. A proposal distribution
$q(\vartheta)$ is constructed near the likely solutions, where
$\vartheta=(\Aacc,\Poms,\theta)$ includes the two noise amplitudes and the four
signal parameters. Independent draws from this proposal receive weights
\begin{equation}
  w_i\propto\frac{\mathcal L(\mathbf d|\vartheta_i)\,\pi(\vartheta_i)}
                    {q(\vartheta_i)},
  \label{eq:importance}
\end{equation}
where $\mathbf d$ is the original full-resolution averaged spectrum and
$\pi(\vartheta)$ is the full parameter prior. The weighted draws approximate
the reference posterior provided the proposal adequately covers its
significant regions.

The proposal is built from searches starting at the SBI posterior mean and
three SBI draws. Each search for a peak of $\lnL$ uses a standard optimiser
for bounded problems (L-BFGS-B), followed by a few Newton steps that use the
second derivatives and are shortened when needed for stability. The noise
amplitudes start at their nominal values. Searches that end within 0.1
Laplace standard deviations of each other are merged, and every peak within
$\Delta\lnL\le10$ of the highest is kept. Each kept peak $j$ has a location
$\mu_j$ and a Laplace covariance $\Sigma_j$, the inverse of its curvature
matrix. Where the peak is too flat in some direction for the curvature to be
measured, the eigenvalues of the curvature matrix are given a minimum value. The
proposal is
\begin{equation}
  q(\vartheta)=0.9\sum_j\eta_j\,t_{\kappa_j}(\vartheta;\mu_j,\gamma_j\Sigma_j)+0.1\,q_{\rm def}(\vartheta),\qquad
  \eta_j\propto e^{\lnL_j}|\Sigma_j|^{1/2},
  \label{eq:proposal}
\end{equation}
a mixture of Student-$t$ distributions $t_\kappa$, which resemble Gaussians with
heavier tails. The number of degrees of freedom $\kappa_j$ sets the tails,
heavier for smaller $\kappa_j$. Here $\gamma_j\Sigma_j$ is the Student-$t$
scale matrix, not its covariance. The latter is
$\kappa_j\gamma_j\Sigma_j/(\kappa_j-2)$ for $\kappa_j>2$. We use
$(\kappa_j,\gamma_j)=(5,1)$ for a well-measured peak and $(3,2)$ for a peak that
needed the minimum curvature. The mixture weights $\eta_j$ are normalised to sum to one. Their
height-times-volume factors approximate the local likelihood mass. The final
posterior probabilities are set by the importance weights, which also include
the prior. The remaining 10\%, $q_{\rm def}$, combines the SBI posterior for
$\theta$ with the conditional Gaussian distribution of
$(\Aacc,\Poms)$ from the Laplace approximation at the highest peak. This component provides additional coverage
beyond the locally fitted peaks. Each round draws 2\,000 candidate sets of all six parameters and weights
them using Eq.~\eqref{eq:importance}. If the highest peak needed the curvature
floor, or a sampled point has $\lnL$ more than 0.01 above that peak, a new
search starts from the best sampled point. The proposal is then rebuilt and
a fresh set of 2\,000 draws is generated. At most two such repair rounds are
allowed, and the final posterior uses the last round's weighted draws. This restart and the
broader treatment of flat peaks were introduced after the development tests
described below.

Pareto-smoothed importance sampling~\cite{Vehtari:2015} fits the tail of the
largest importance weights and replaces them with smoothed values. Its
estimated tail index $\khat$ is used to identify unstable weights, and values
above 0.7 flag the result as unreliable. Such behaviour can arise when the
proposal poorly covers a significant region, including a competing peak.
The $\khat$ check cannot guarantee detection of a peak that receives no draws.
All posterior summaries use the smoothed weights.

In the initial calibration test on 100 development simulations, the guided
procedure passed 11 of 12 checks. It fell one simulation short of the 95\%
coverage requirement for $\OExt$. Neither a modulation-corrected likelihood nor a
wider proposal resolved the failure. Inspection identified two searches that
stopped before a higher peak and two cases with poorly determined curvature.
These cases motivated the restart and flat-peak rules above. The revised
procedure was then fixed before the final test on 100 new simulations
(Sec.~\ref{sec:reliability}).

As a preliminary efficiency check, we compared the peak-based proposal with
direct sampling from the SBI posterior. This comparison was performed before
we added the restart rule and revised the treatment of flat likelihood peaks.
Effective sample size measures how many equally weighted samples would give
comparable precision. For the proposal built around refined likelihood
peaks, its median was 0.156 times the number of draws over 60 development
observations. The SBI-only proposal gave ratios of $7\times10^{-4}$ to
$2\times10^{-2}$ on three observations from this set. This limited check
suggests an efficiency gain of roughly 10--200, although it is not a systematic
comparison of the final procedure with alternative proposals. All three
searches that stopped far below a peak in that development study were flagged
by $\khat$.

\subsection{Calibration criteria}

The calibration tests of Sec.~\ref{sec:reliability} use the probability
integral transform (PIT), the posterior probability below the true value of a
parameter. For a calibrated method the PIT values of many simulations are
uniformly distributed between 0 and 1. SBC~\cite{Talts:2018} uses the equivalent
rank of the truth among posterior draws, and TARP~\cite{Lemos:2023} tests the
four signal parameters jointly. The criteria for SBI, fixed before any SBI
simulation was generated, are an SBC $\chi^2$ test per parameter with
$p>0.0025$, 68\% and 95\% coverage within three binomial standard errors
($\pm7.0$ and $\pm3.3$ percentage points for 400 simulations), and TARP with
$p>0.01$. A calibrated method fails these by chance about 4\% of the time. The
criteria for SBI-guided inference are a Kolmogorov--Smirnov test of the PIT
values with $p>0.005$, and 68\% and 95\% counts between 55 and 81 and between 89
and 100 out of 100, which a calibrated method fails by chance 1.3--4.8\% of the
time. Table~\ref{tab:calib} gives the numerical results,
and Sec.~\ref{sec:reliability} discusses the larger-set SBI discrepancy and the
flagged guided outputs. In the 2\,000-observation development set, the affected
Galactic-amplitude posterior is drawn towards the prior centre. Four further
training initialisations show a small SBC error in different parameters.
For the final 100 guided-test observations, twelve required a repeated search
and 44 used the broader proposal for at least one flat peak.

\begin{table}[tbp]
\centering
\caption{Calibration results. For SBI on its 400 test simulations, the 68\% and
95\% coverage in per cent and the SBC $p$-value (TARP $p=0.67$). For SBI-guided
exact inference on the 100 blind-test simulations, the 68\% and 95\% counts and
the KS $p$-value of the PIT values. All checks pass.}
\label{tab:calib}
\begin{tabular}{llcccc}
\toprule
 & & $\OGal$ & $\OExt$ & $\Ob$ & $\fb$ \\
\midrule
SBI & 68\% coverage (\%) & 68.0 & 64.2 & 72.2 & 67.5 \\
 & 95\% coverage (\%) & 93.0 & 91.8 & 94.0 & 92.0 \\
 & SBC $p$ & 0.059 & 0.154 & 0.476 & 0.365 \\
\midrule
SBI-guided & 68\% coverage (of 100) & 61 & 61 & 67 & 66 \\
 & 95\% coverage (of 100) & 95 & 92 & 92 & 91 \\
 & PIT KS $p$ & 0.183 & 0.556 & 0.386 & 0.813 \\
\bottomrule
\end{tabular}
\end{table}

\subsection{Probabilities of competing solutions}

Table~\ref{tab:regions} supplements the comparison in
Sec.~\ref{sec:benchmark}. It includes a repeat with 20\,000 importance samples
for the same three-solution observation, in addition to the standard
2\,000-sample calculation.

\begin{table}[tbp]

\centering
\caption{Probability of the three regions for the three-peak test observation
(Fig.~\ref{fig:corner}). Each sample is assigned to the nearest component of a
three-component Gaussian mixture fitted to the MCMC samples in the four signal
parameters (Mahalanobis distance). Guided probabilities include the importance weights. The quoted errors
are estimated from batches of MCMC samples (a lower bound), bootstrap
resampling of the weighted guided samples, and binomial sampling errors for
50\,000 draws from each neural estimator. The neural-estimator errors exclude
variation from retraining.}
\label{tab:regions}
\small
\begin{tabular}{lccc}
\toprule
Method & Region 1 & Region 2 & Region 3 \\
\midrule
Exact-likelihood MCMC & $53.2\pm1.0$\% & $35.3\pm1.1$\% & $11.5\pm0.6$\% \\
SBI-guided, 2\,000 draws & $52.8\pm2.1$\% & $34.9\pm2.0$\% & $12.2\pm1.2$\% \\
SBI-guided, 20\,000 draws & $51.7\pm0.8$\% & $36.4\pm0.7$\% & $12.0\pm0.5$\% \\
SBI-S & $49.8\pm0.2$\% & $16.6\pm0.2$\% & $33.7\pm0.2$\% \\
TSNPE-S & $26.8\pm0.2$\% & $19.2\pm0.2$\% & $53.9\pm0.2$\% \\
TSNPE-R (longer patience) & $21.4\pm0.2$\% & $17.1\pm0.2$\% & $61.5\pm0.2$\% \\
\bottomrule
\end{tabular}
\end{table}

\section{Fisher analysis and additional parameter points}\label{app:fisher}

A Fisher-information calculation estimates the local precision with which
parameters can be measured near a specified injected point. In the Gaussian
approximation, the inverse of the data Fisher matrix $F_{\rm data}$ gives the
parameter covariance, $C_{\rm data}=F_{\rm data}^{-1}$. The marginal standard
deviation of parameter $i$ is $\sqrt{(C_{\rm data})_{ii}}$, with the other
parameters allowed to vary. Here each signal-parameter width is marginalised over the
other five parameters, including $\Aacc$ and $\Poms$, and compared with the
corresponding width from the exact likelihood of Eq.~\eqref{eq:lnL}. This is a
local comparison and does not describe separated posterior peaks or
non-Gaussian tails. As a check on frequency binning alone, we replace the full grid by the
network's 512 bins and approximate each binned PSD by a Gamma distribution
with the same mean and variance. In this calculation, the Fisher widths
change by at most 0.7\%.

We call a Fisher comparison \emph{prior-free} when no information from the
analysis prior is added to $F_{\rm data}$ before inversion. Its widths measure
local information supplied by the data representation alone, without
restricting the Gaussian uncertainty region to the finite analysis-prior
ranges. This terminology refers only to the uncertainty calculation. The
summaries are still trained on simulations drawn from their stated training
distributions.

Figure~\ref{fig:fisher} instead includes a Gaussian approximation to the
analysis prior on the four signal parameters $\theta$. For a uniform range
$\theta_i\in[a_i,b_i]$ in Table~\ref{tab:priors}, we use a Gaussian with the same
variance, $\sigma_{{\rm prior},i}^2=(b_i-a_i)^2/12$. Its contribution is added
before marginalisation,
\begin{equation}
  F_{\rm fig}=F_{\rm data}+P,\qquad
  P_{ij}=\delta_{ij}\frac{12}{(b_i-a_i)^2}
  \quad (i,j\text{ in the signal-parameter block}),
  \label{eq:fisherprior}
\end{equation}
with all entries involving the two instrument-noise parameters set to zero.
The same $P$ is used for the exact likelihood and both summaries. The plotted
covariance is $F_{\rm fig}^{-1}$, and its marginal two-dimensional 68\% contours
use $\Delta\chi^2=2.30$. Matching the variance is an approximation to the finite
extent of the uniform prior, rather than an exact treatment of its hard
boundaries. Thus the figure includes prior information and is not a
prior-free comparison.

For a summary statistic $s$, the data Fisher matrix used here is
$G^{\mathsf T}\Sigma^{-1}G$, where $G$ is the derivative of the mean of $s$ with
respect to the parameters and $\Sigma$ its covariance. Both are measured from
simulations, $\Sigma$ from 600 noise realisations at the true parameters and
$G$ from central differences with 240 realisations on each side, using the
same noise realisations on both sides. The inverse covariance is corrected for
the bias of a covariance estimated from a finite sample, and the expected
contribution of noise in the numerical derivatives is subtracted. This expression uses changes in the mean summary and holds its covariance
fixed at the injected parameters. With exact moments and derivatives, it is
a lower bound on the summary's Fisher information, since other changes in its
distribution can also carry information. The corresponding width ratios
therefore estimate an upper bound on the local loss of precision. Finite
simulation samples and derivative errors limit the numerical accuracy of
this bound.

We supplement the benchmark with nine parameter sets selected from the
first 100 test observations. Eligible cases have peak searches that agree to
within 0.1 Laplace standard deviations in every parameter and a guided
reference with $\khat<0.7$ (App.~\ref{app:guided}). To spread the selection across the analysis range,
we map the four injected signal parameters to the unit prior box, start with
the eligible case closest to its centre, and repeatedly select the case
farthest from those already chosen, using its distance to the nearest
selected case. Each selected case retains its own noise amplitudes and
modulation.

Table~\ref{tab:fisherpoints} lists the ten cases and their prior-free width
ratios. Nine cases change by at most 3\% when the derivative step is halved
between the two smallest values. The remaining case, marked in the table,
has not fully converged. Its separated-summary widths were still increasing as the
step was reduced, and the converged widths may therefore be higher than those listed.
These are local comparisons at selected parameter points, not a test over
the full prior or over observations with multiple posterior solutions.

For each signal parameter, we take the median separated-to-exact width
ratio over the ten cases. These four medians range from 1.6 to 2.2
(Table~\ref{tab:fisherpoints}). They are not bounds on individual observations.

{
We also divide the SBI-S central 68\% interval half-width by the
separated-summary Fisher standard deviation for each case and parameter.
The four parameter-wise median ratios are 2.2--3.2. This comparison suggests
additional loss of precision in the posterior estimator, subject to the
local Fisher approximation and the case selection above.
\par}

\begin{table}[tbp]
\centering
\begin{threeparttable}
\caption{Prior-free Fisher widths at the ten parameter points. The first four
columns give the injected signal parameters, as $\log_{10}$ of the quantities
named, with $\fb$ in Hz. The remaining columns give the marginal Fisher
standard deviation of each signal parameter from the separated and raw
32-component summaries, divided by that from the exact likelihood, without
the analysis prior. The last row gives the medians over the ten points.}
\label{tab:fisherpoints}
\footnotesize
\setlength{\tabcolsep}{3.2pt}
\begin{tabular}{lrrrr@{\hspace{8pt}}rrrr@{\hspace{8pt}}rrrr}
\toprule
 & \multicolumn{4}{c}{Injected values} & \multicolumn{4}{c}{Separated / exact} & \multicolumn{4}{c}{Raw / exact} \\
\cmidrule(lr){2-5}\cmidrule(lr){6-9}\cmidrule(lr){10-13}
Point & $\OGal$ & $\OExt$ & $\Ob$ & $\fb$ & $\OGal$ & $\OExt$ & $\Ob$ & $\fb$ & $\OGal$ & $\OExt$ & $\Ob$ & $\fb$ \\
\midrule
1\tnote{a} & $-$7.84 & $-$12.38 & $-$11.50 & $-$3.00 & 2.90 & 2.02 & 2.53 & 2.53 & 76 & 37 & 70 & 69 \\
2 & $-$7.31 & $-$12.67 & $-$11.71 & $-$3.48 & 3.43 & 1.60 & 2.05 & 1.79 & 4.4 & 1.5 & 3.4 & 2.9 \\
3\tnote{b} & $-$7.85 & $-$12.08 & $-$11.66 & $-$3.12 & 2.48 & 1.69 & 2.07 & 2.12 & 18 & 3.3 & 15 & 16 \\
4 & $-$8.13 & $-$12.89 & $-$12.15 & $-$2.84 & 1.69 & 1.54 & 1.54 & 1.53 & 82 & 57 & 93 & 85 \\
5 & $-$8.33 & $-$12.81 & $-$11.00 & $-$3.07 & 2.23 & 1.81 & 2.10 & 2.15 & 27 & 5.8 & 23 & 24 \\
6 & $-$8.33 & $-$11.87 & $-$11.49 & $-$2.42 & 1.62 & 1.58 & 1.52 & 1.43 & 57 & 84 & 81 & 100 \\
7 & $-$8.47 & $-$12.47 & $-$10.54 & $-$3.14 & 2.39 & 1.55 & 2.10 & 2.01 & 16 & 2.0 & 14 & 15 \\
8 & $-$8.44 & $-$11.93 & $-$11.74 & $-$3.55 & 1.69 & 1.91 & 1.59 & 1.55 & 5.0 & 1.6 & 4.7 & 2.1 \\
9 & $-$8.19 & $-$12.38 & $-$11.09 & $-$2.13 & 1.80 & 1.54 & 1.49 & 1.40 & 43 & 80 & 79 & 12 \\
10 & $-$7.07 & $-$12.61 & $-$11.37 & $-$2.82 & 2.08 & 1.55 & 1.91 & 1.74 & 38 & 57 & 73 & 74 \\
\midrule
Median & & & & & 2.16 & 1.59 & 1.98 & 1.77 & 33 & 22 & 47 & 20 \\
\bottomrule
\end{tabular}
\begin{tablenotes}\footnotesize
\item[a] The benchmark of Fig.~\ref{fig:fisher}.
\item[b] The derivatives have not fully converged, and the separated widths may be higher than those listed.
\end{tablenotes}
\end{threeparttable}
\end{table}

The raw summary statistic is the network input, standardised bin by bin over
the SBI training simulations and projected onto its $K$ leading principal
components. The number of components needed to match the separated 32 was
found with a faster calculation that assumes independent Gamma scatter in each
bin. For up to 256 raw components, its widths agree with the simulation-based
calculation to within a few per cent in the median comparison.
For each case and signal parameter, we find the smallest number of raw
principal components whose Fisher width is no larger than that of the
separated 32-component summary. Taking the median over the ten cases gives
256--320 components, depending on the parameter. The upper end of this range
therefore relies on the analytic shortcut alone. This is the basis of the
"about 300" comparison in the main text. Without the analysis
prior, 32 raw components barely constrain a near-degenerate combination of
$\OGal$, $\Ob$ and $\fb$, and the raw widths are 37--76 times the exact ones at
the benchmark. This is why Fig.~\ref{fig:fisher} includes the Gaussian prior
contribution of Eq.~\eqref{eq:fisherprior}, which limits that weakly constrained
direction. Table~\ref{tab:fisherpoints} omits this contribution so that the
comparison isolates information supplied by the data representations.

\bibliographystyle{JHEP}
\bibliography{refs}
\end{document}